# Three-Dimensional Solid-State Lithium-Ion Batteries Fabricated Via Conformal Vapor-Phase Chemistry

*Alexander J. Pearse[a], Thomas E. Schmitt[a], Emily Sahadeo[b], David Stewart[a,f], Alexander C. Kozen[d], Konstantinos Gerasopoulos[c], A. Alec Talin[e], Sang Bok Lee[b], Gary Rubloff[a,f], Keith E. Gregorczyk[a]*

[a]Department of Materials Science and Engineering, University of Maryland, College Park, MD 20740

[b]Department of Chemistry, University of Maryland, College Park, MD 20742

[c]Research and Exploratory Development Department, The Johns Hopkins University Applied Physics Laboratory, Laurel, MD 20723

[d]American Society for Engineering Education, residing at the U.S. Naval Research Laboratory 1818 N St NW, Suite 600 Washington DC, 20036

[e]Sandia National Laboratories, Livermore, CA 94551

[f]Institute for Systems Research, University of Maryland, College Park, MD 20742

**ABSTRACT:**

Thin film solid state lithium-based batteries (TSSBs) are increasingly attractive for their intrinsic safety due to the use of a nonflammable solid electrolyte, cycling stability, and ability to be easily patterned in small form factors. However, existing methods for fabricating TSSBs are limited to planar geometries, which severely limits areal energy density when the electrodes are kept sufficiently thin to achieve high areal power. In order to circumvent this limitation, we report the first successful fabrication of fully conformal, 3D full cell TSSBs formed in micromachined silicon substrates with aspect ratios up to ~10 using atomic layer deposition (ALD) at low processing temperatures (≤ 250C) to deposit all active battery components. The cells utilize a prelithiated $LiV_2O_5$ cathode, a very thin (40 – 100 nm) LiPON-like lithium polyphosphazene ($Li_2PO_2N$) solid electrolyte, and a $SnN_x$ conversion anode, along with Ru and TiN current collectors. Planar all-ALD solid state cells deliver 37 µAh/cm$^2$·µm normalized to the cathode thickness with only 0.02% per-cycle capacity loss for hundreds of cycles. Fabrication of full cells in 3D substrates increases the areal discharge capacity by up to a factor of 9.3x while simultaneously improving the rate performance, which corresponds well to trends identified by finite element simulations of the cathode film. This work shows that the exceptional conformality of ALD, combined with conventional semiconductor fabrication methods, provides an avenue for the successful realization of long-sought 3D TSSBs which provide power performance scaling in regimes inaccessible to planar form factor devices.

# 1. Motivation
## 1.1 Thin Film Solid State Batteries

State-of-the-art lithium-ion batteries, which utilize particle-based composite electrodes and organic liquid electrolytes, represent an enabling and successful technology. However, conventional battery systems suffer from safety concerns relating to the combustion of the flammable electrolyte, are difficult to implement in nonstandard or very small form factors, and exhibit limited high-power performance due to sluggish transport in the stochastically formed particulate electrodes.[1,2] As a result, alternative energy storage technologies are needed for applications which require very small, intrinsically safe, or high power sources, such as distributed sensor networks, implantable medical devices, smart cards, and microelectromechanical systems.[3–5] These applications all require high volumetric energy density power sources, and in particular require sources with high *areal* energy and power density in order for the size of the integrated battery to not dominate the entire device.[6]

Thin-film solid state batteries (TSSBs), which have been under development for several decades and have been commercialized, promise to solve some of the issues with conventional Li-ion cells.[4,7,8] TSSBs are fabricated using traditional semiconductor manufacturing methods, and replace the liquid electrolyte with an inert lithium-conducting glass- typically a member of the lithium phosphorus oxynitride (LiPON) family.[9] The use of a solid electrolyte renders the batteries nonflammable and tolerant to high temperatures, and the additional chemomechanical stability imparted by the use of LiPON allows TSSBs to operate for thousands of cycles. TSSBs are also geometrically optimal, in the sense that they are effectively one-dimensional structures which minimize inhomogeneous internal current distributions and require ionic and electronic transport length scales on the order of a few micrometers at most.[10] In addition, using mature semiconductor manufacturing methods allows for the straightforward fabrication of batteries with sub-mm lateral dimensions, and the solid electrolyte can readily have submicron thickness to minimize internal resistance.[11,12]

However, the use of nonconformal, line-of-sight physical vapor deposition (PVD) to deposit the battery components critically limits the form factor of TSSBs to planar substrates, which in turn limits TSSBs to capacities on the order of 0.1 mAh/cm$^2$.[13] On a planar substrate, the cathode cannot easily be made more than ~5-10 μm thick before film stress results in delamination and cracking.[4] In addition, in a planar geometry, there is a fundamental tradeoff between the deliverable areal power density and energy density- the only method available to PVD to improve the energy density is to increase the cathode thickness, but a thicker electrode cannot deliver its full capacity at high power due to the increased diffusion length for Li ions.[14] As an illustration of the problem, replacing the 1960 mAh standard battery in an iPhone 7 with a state-of-the-art commercial LiCoO$_2$/LiPON/Li TSSB as made by current fabrication techniques would require a ~1.3 m$^2$ cell.[7]

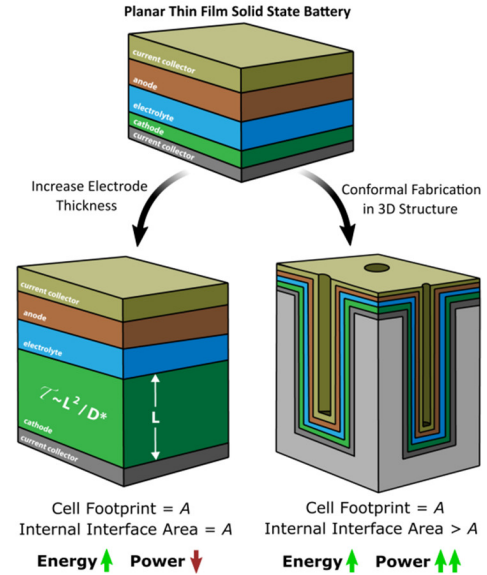

**Figure 1:** Methods of improving energy storage metrics for thin film solid state batteries. Increasing the electrode thickness scales energy density at the expense of power density due to the rapid increase in the characteristic diffusion time for ions in the thick electrode. Fabricating TSSBs in a 3D structure both increases areal material loading and increases power performance through a decrease in the internal current density due to the larger internal surface area.

## 1.2 Moving to 3D Architectures

While battery performance can be incrementally improved through materials development, i.e. through the use of higher capacity/voltage electrodes, enabling TSSB fabrication on 3D substrates, analogous to the strategies now employed to improve transistor and NAND flash memory areal packing density in semiconductor devices, would be a more fundamentally enabling approach. By simply increasing the surface area upon which a TSSB is grown, the areal capacity can be improved while maintaining optimal (usually nanoscale) electrode thicknesses and locally 1-D current distributions for high specific power densities (Figure 1). 3D TSSBs were first proposed in detail in 2004 by Long et al. and have been a goal of the energy storage community since at least that time, with several subsequent reports outlining the benefits of the architecture. [3,14–17]

While conceptually simple, the experimental realization of such 3D TSSBs has been stymied by the need for extremely conformal deposition techniques for the active layers of a battery, and only a handful of examples of working cells have been reported. Arguably the most successful design was described by Peled et al., who developed Li-ion full cells formed in microchannel plates using a mixture of electrodeposition and physical impregnation steps.[18] While the device performance was impressive, this method may be difficult to scale down and does not achieve ideal layer self-alignment. Talin et al. have shown the 3D TSSBs can be fabricated over 3D scaffolds of modest aspect ratio using RF sputtering, but in two reports such cells actually performed worse than planar analogues due to electrical leakage and inhomogeneous current distributions, again because the active layers did not have a uniform thickness.[13,19]

**1.3 Conformal Deposition for 3D TSSBs**

A better approach to fabricating full cells with uniform thickness in high aspect ratio 3D structures is to use vapor-phase chemistry methods such as chemical metalorganic chemical vapor deposition (CVD) or atomic layer deposition (ALD). ALD in particular is capable of uniform growth in structures with aspect ratios in the hundreds, and works at temperatures low enough (generally below 300C) to enable deposition on flexible polymeric substrates.[20,21] Importantly, ALD and CVD are both mature techniques integrated with existing semiconductor manufacturing.

As of yet, there are no published examples of full solid state batteries in which all active components are grown via a conformal vapor-phase deposition technique such as CVD or ALD, despite significant work exploring the growth and electrochemical performance of individual battery components in isolation.[22–24] Growing thin film materials with vapor-phase chemistry is generally more complicated than by PVD, as each process involves a carefully designed surface-mediated chemical reaction in contrast to ablating and re-depositing material from a prefabricated target. This leads to a limited selection of high-quality electrode materials, discussed further below. Conformal inorganic solid electrolytes are even more limited, and include CVD lithium phosphate, ALD LiPON variants, and a few mixed lithium metal oxides.[21,25–29] ALD-grown electrolytes are particularly attractive because of their unique ability to realize complete electrical isolation in full cells at thicknesses below 50nm, reducing cell size, resistance, and fabrication time.[21] An additional complication for growing batteries with vapor-phase deposition arises from the need for each layer to be stable in the deposition conditions of subsequent layers, and each electrode/electrolyte interface must not be damaged by incoming reactive species, as can sometimes occur with plasma radicals or strongly reducing/oxidizing precursors. Finally, the exceptional conformality of ALD/CVD becomes a double-edged sword when it comes to patterning multi-layer active devices such as full batteries. While PVD-grown TSSBs can be patterned via simple shadow-masking lithography, more conformal techniques require subtractive approaches in order to define batteries of a specific size without electrically shorting the anode and cathode.

**1.4 Conceptual Overview**

Here, we will describe the successful fabrication of a 3D solid state thin film Li-ion battery comprised of 5 conformal layers. The electrochemical couple is formed by a $LiV_2O_5$ cathode and a $Sn_xN$ anode, each paired with a current collector (Ru and TiN, respectively). For the solid state electrolyte, we utilize a previously described ALD lithium polyphosphazene (LPZ) film, which is a polymorph of LiPON.[21] We take a stepwise approach to measuring intrinsic materials properties, mutual materials compatibility, and the effects of cell geometry on performance.

- First, we construct planar solid state half-cells comprised of either the anode or cathode paired with LPZ and a thin film Li metal anode in order to assess operating potential range and electrochemical kinetics of each electrode, as well as the mutual compatibility of the electrode and ALD electrolyte.
- We then develop a finite-element simulation of the cathode based on half-cell data in order to predict performance scaling for planar vs. 3D devices.
- Next, we fabricate planar full cells using vapor-phase grown materials in order to assess the kinetics and stability of the full cell chemistry, as well as optimal capacity matching for the electrodes. Optimized all-ALD full cells exhibit excellent cycling stability and reach 37 µAh/cm$^2$·µm normalized to the cathode thickness.
- Finally, we successfully integrate the full cell film stack with 3D-structured substrates and demonstrate the simultaneous upscaling of both areal capacity and rate performance. Demonstrated benefits of 3D structuring include an order-of-magnitude improvement in areal capacity, improved rate performance, and improved round-trip efficiency.

## 2. Results and Discussion
### 2.1 Electrode Selection

Conformal battery materials were primarily chosen based on three criteria: (1) ability to be synthesized at moderate or low temperatures (≤ 250C) in the active phase to enable growth on a broad variety of substrates, (2) minimal complexity of fabrication to reduce production time (i.e. avoiding more than two precursors per ALD process), and (3) mutual compatibility with regard to both synthesis conditions and electrochemical stability.

The selection of developed cathode and anode materials available for growth via ALD is limited because of the difficulty in growing crystalline, Li-containing multicomponent oxides without a high-temperature annealing step. In the case of multicomponent oxides, it has been reported that common Li precursors (including LiO$^t$Bu) frequently do not exhibit self-limiting growth on oxide surfaces due to their tendency to directly reduce metal ions as a side-reaction,[30] analogous to direct chemical lithiation. As a result, controlling the Li stoichiometry in ALD-grown oxide films is challenging, and all reported Li-containing ALD-grown crystalline cathodes require a high-temperature annealing step. Materials reported to be grown by ALD in this manner include LiCoO$_2$, LiFePO$_4$, and LiMn$_2$O$_4$, reaching varying levels of performance relative to bulk synthesis methods as a result of impurity inclusion from nonstoichiometric Li incorporation.[30–32] Another disadvantage to growing cathode films via multicomponent ALD processes is the increase in deposition time; adding an additional Li-incorporating subcycle to a binary ALD process roughly halves the growth rate.

**Cathode:** To avoid these issues, we have taken a simpler approach to synthesizing a conformal prelithiated cathode. The ALD reaction between VO(OC$_3$H$_7$)$_3$ (VTOP) and O$_3$ produces crystalline V$_2$O$_5$ at 170C.[33] We then employ electrochemical lithiation as a conformal technique to rapidly transform the deposited V$_2$O$_5$ through the reaction V$_2$O$_5$ + Li$^+$ + e$^-$ → LiV$_2$O$_5$ in a LiClO$_4$/propylene carbonate electrolyte, which can then be incorporated into a full solid state battery. At 3.4 V vs Li/Li$^+$, the potential of the lithium intercalation reaction is within the electrochemical stability window of the electrolyte, resulting in minimal surface contamination.[34] Orthorhombic V$_2$O$_5$ is a commonly employed TSSB cathode material exhibiting multiple lithium-intercalating phase transitions, and when cycled in the one Li per unit cell electrochemical window, exhibits an acceptable capacity (49 µAh/cm$^2$ µm), high voltage (> 3V vs. Li/Li$^+$), and excellent cycling stability.[35,36] The primary downside of the material is a limited rate capability due to a relatively low average chemical diffusion coefficient for Li (≈ 10$^{-13}$ cm$^2$/s),[33,37] although as we will show this problem can be partially mitigated through the use of nanoscale films and 3D structuring.

**Anode:** On the anode side, the use of Li metal is currently ruled out for lack of a plausible ALD process. Thus, any all-ALD conformal SSB will be a Li-ion cell. ALD processes for single-element alloying type anodes such as Si, Sn or Al are also yet to be developed, although relatively low temperature and conformal CVD processes are available for Si in particular.[38] Conversion-type anodes, which undergo a first-cycle irreversible transformation into an active phase, are a promising alternative and are readily made via vapor phase chemistry. For the batteries in this paper, we utilize a novel ALD process for amorphous tin nitride (SnN$_x$) grown using tetrakisdimethylamidotin (TDMASn) and a N$_2$ plasma, which will be described fully in a separate publication. This process produces an amorphous material with a temperature-dependent composition of approx. SnN$_x$ ($x$ ≈ 2 at 200C) plus a small amount of carbon and oxygen incorporation. Tin nitride, which has been utilized previously in lithium-ion configuration planar thin film SSBs,[4] was chosen for its low electrochemical potential (operating below 1V vs. Li/Li$^+$) and use of nitrogen as the oxidant, which was expected to be less damaging to the electrolyte/anode interface than H$_2$O.[39]

**Current Collectors:** A conformal battery also requires conformal current collectors in intimate contact with both electrodes. In this work, we utilize ALD Ru as the cathode current collector and ALD TiN as the anode collector, which has been previously shown to be a good conductive Li diffusion barrier.[40]

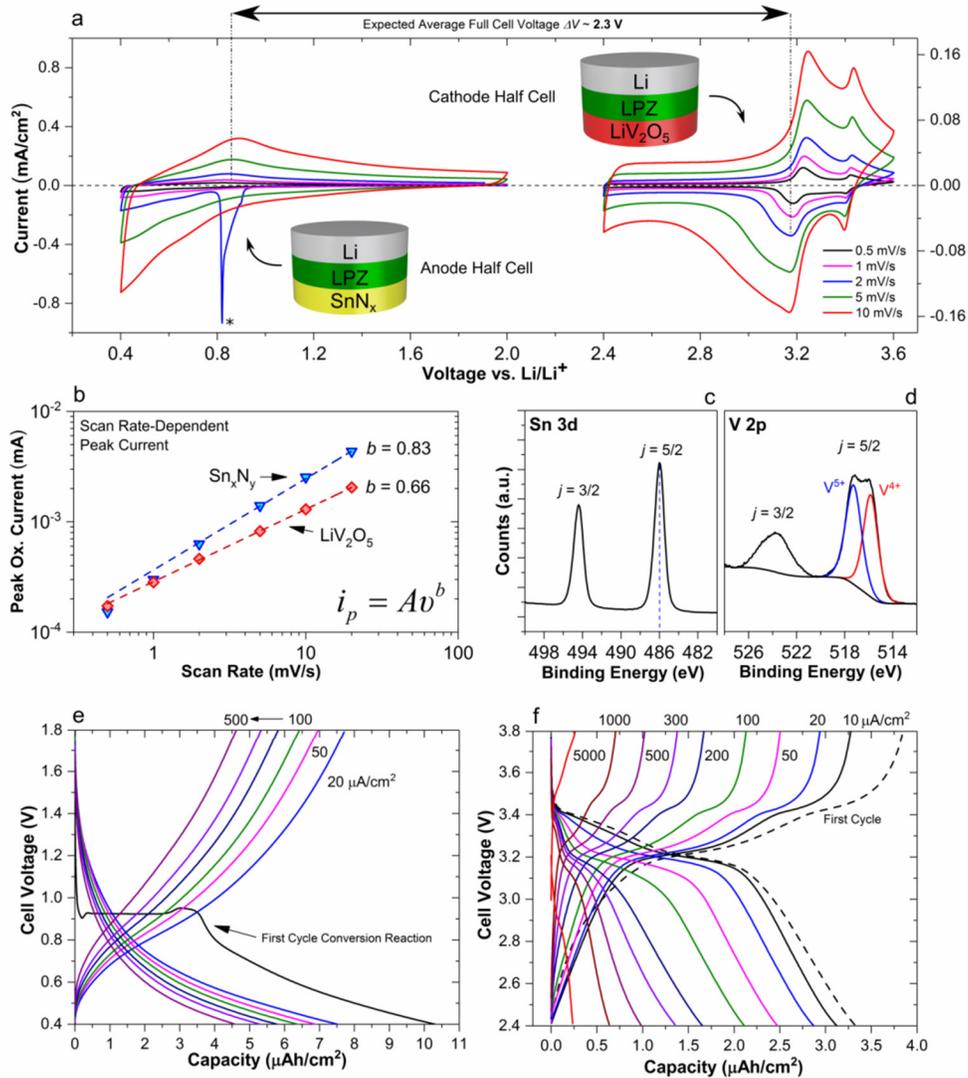

**Figure 2:** Electrochemical and XPS characterization of thin film LiV$_2$O$_5$ and SnN$_x$ electrodes. (a) Cyclic voltammetry at different scan rates of planar solid state half cells. Electrodes were tested in the two-electrode configuration Pt/electrode material/80nm LPZ/3000nm Li, using 25nm SnN$_x$ or 70nm LiV$_2$O$_5$. The first-cycle conversion reaction for SnN$_x$ is indicated by (*). (b) Log-log plot of the peak oxidative current as a function of scan rate for the half-cells. Data are fit to a power law. (c) High-resolution XPS scan of Sn 3d region of as-deposited SnN$_x$, indicating the initial oxidized state of Sn. (d) High-resolution XPS scan of the V 2p region of as-made LiV$_2$O$_5$ with fitting, indicating the presence of V$^{4+}$ as expected for the discharged state. (e,f) Galvanostatic rate performance of solid state half-cells with a configuration identical to those in (a), tested with current densities between 20 and 500 μA/cm$^2$ for SnN$_x$ and between 10 and 5000 μA/cm$^2$ for LiV$_2$O$_5$.

## 2.2 Solid State Half-Cell Characterization

Figure 2 characterizes the electrochemical performance and initial composition of thin films of the LiV$_2$O$_5$ cathode (70 nm) and the SnN$_x$ anode (25 nm), tested in an all-solid-state half-cell configuration via coating with approx. 80nm of ALD lithium polyphosphazene (LPZ) as the solid electrolyte followed by thermal evaporation of 3 μm of metallic Li as the anode. This configuration allows for the determination of the capacity, kinetics, and electrochemical potential of the electrode films, as the Li anode acts as an infinite Li source and a relatively reliable reference electrode even in a two-electrode configuration. We have previously shown the ALD LPZ/Li interface to be stable.[21] Figure 2a shows cyclic voltammetry (CV) of the anode and cathode films at scan rates between 0.5 and 10 mV/s plotted on a single axis referenced to Li/Li$^+$ at 0V. Both half-cells show repeatable anodic and cathodic processes associated with the storage of Li ions, indicating that the LPZ ALD process is chemically compatible with each material.

**SnN$_x$:** Similar to other conversion-type nitrides, the SnN$_x$ film undergoes a first-cycle conversion reaction, indicated by the sharp asymmetric peak located between 0.9 and 0.8V vs. Li/Li+ (Figure 1a, left), of the general form:

$$\text{SnN}_x + 3x\text{Li}^+ + 3xe^- \rightarrow x\text{Li}_3\text{N} + \text{Sn}^0$$

X-ray photoelectron spectroscopy (XPS) of the Sn 3d $j$=5/2 core level at 486 eV in as-grown SnN$_x$ films (Figure 2c) suggests the initial average valence state of Sn is close to +4, leading to an irreversible capacity loss associated with reduction of Sn(IV) to Sn(0).[41] Following the formation of tin nanocrystals embedded in a Li$_3$N matrix, Li begins to directly alloy with Sn, forming a series of metallic Li-Sn compounds. We limit the lower potential of the SnN$_x$ films to 0.4 V vs. Li, which should in principle correspond to the formation of the LiSn phase.[42] Further lithiating the SnN$_x$ films lead to occasional cell failure through the formation of soft electrical shorts, the origin of which remains under investigation. On the reverse scan, and in contrast with ALD SnO$_2$ anodes which exhibit multiple delithiation peaks (Figure S1), the Li dealloying process occurs in a single peak located at approx. 0.9 V vs. Li/Li+ which helps to maintain a higher discharge voltage when utilized as the anode in a full cell.

**LiV$_2$O$_5$:** Cyclic voltammetry of the LiV$_2$O$_5$ cathode films (Figure 2a, right) between 2.4 and 3.6V vs. Li/Li+ reveals the its characteristic doublet, removing or adding 1 Li per formula unit in two steps of approximately 0.5 Li per peak at 3.4 V (ε – α transition) and 3.2 V (δ – ε transition).[43] Importantly for use in a Li-ion configuration, the initial charging sweep also reveals these characteristic peaks, indicating the lithium inserted during the prelithiation process is fully active after cell fabrication. Component analysis of XPS of the V 2p $j$ = 5/2 core level of the as-made LiV$_2$O$_5$ (Figure 2d) shows an equal population of V$^{5+}$ and V$^{4+}$, confirming the successful formation of the desired phase.

**Analysis and Kinetics:** Combining information from the anode and cathode half-cells affords important predictions about the full cell. Based on the position of the delithation peak of the anode and the lithiation peaks of the cathode, an average capacity-matched LiV$_2$O$_5$/SnN$_x$ full cell discharge potential can be estimated to be approximately 2.3V. In addition, we are able to decouple anode and cathode kinetics and identify the rate limiting step in a full cell, under the assumption that the Li/Li$^+$ couple at the Li/LPZ interface is facile. Figure 1b plots the peak oxidative current $i_p$ as a function of CV scan rate υ for both materials. It is well known that reducing the thickness $t$ of battery electrode materials enhances rate performance through several mechanisms, including a simple reduction in characteristic diffusion time $\tau \propto t^2/D^*$ as well as the increased prominence of surface and near-surface pseudocapacitive charge storage mechanisms. Fitting $i_p$ to the power law $i_p(v) = Av^b$ provides insight into the nature of charge storage in thin films, with $b = 1$ corresponding to purely capacitive storage, $b = 0.5$ corresponding to diffusively-limited charge storage, and intermediate values corresponding to a combination of these effects.[43,44] We find $b = 0.66$ for the LiV$_2$O$_5$ film and $b = 0.83$ for the SnN$_x$ film after conversion, indicating a considerable contribution of non-diffusion limited storage in the anode in particular and that transport in a full cell utilizing this electrode pair will be rate-limited by diffusion in the cathode.

Galvanostatic rate testing, shown in Figure 2e and 2f, supports the preceding observations. The SnN$_x$ anode half-cells maintain 60% of their capacity between 20 and 500 μA/cm$^2$ cell current compared with 37% retention for LiV$_2$O$_5$ in the same interval. While some of this difference can be attributed to the fact that the tested anode film is thinner than the cathode (25 nm vs. 70 nm), we note that at the same current density, the capacity of the SnN$_x$ film with a 0.4V cutoff potential (300 μAh/cm$^2$·μm @ 20 μA/cm$^2$) is dramatically higher than LiV$_2$O$_5$ (38.6 μAh/cm$^2$·μm @ 20 μA/cm$^2$). Thus, in a capacity-matched full cell, the anode film will always be ~8x thinner, and comparing the rate performance of thin anodes to thicker cathodes is a device-relevant regime. Further decreasing the testing current for LiV$_2$O$_5$ to 10 μA/cm$^2$ yields a discharge capacity of 44.4 μAh/cm$^2$·μm, which is 90.6% of the theoretical capacity.

### 2.3 Simulation of Performance Scaling of LiV$_2$O$_5$: Planar vs. 3D

In this section, we briefly develop a one-dimensional finite-element simulation of the LiV$_2$O$_5$ cathode coupled with the ALD LPZ electrolyte in order to illustrate trends in performance scaling, given that diffusion in the cathode is the rate limiting process. We model Li transport in the LPZ electrolyte using the Nernst-Planck equation based on the work by Danilov et al., and charge transfer at the electrode/electrolyte interface with Butler-Volmer kinetics.[13,45] Lithium transport in the cathode film is modelled using Fick's law $\frac{dc_{Li}}{dt} = \frac{d}{dx}\left[D^*(x, c_{Li})\frac{dc_{Li}}{dx}\right]$ where the Li chemical diffusion coefficient $D^*$ can in principle have a positional and concentration dependence. Full details of the model can be found in Supplementary Discussion 1 and Table S1. Figure 2a shows simulated discharge curves for a 70nm $LiV_2O_5$ film at the same current densities tested in Figure 2f. The optimized model adequately captures the both the experimental overpotentials and trend in discharge capacity (plotted in the inset). The model confirms that the primary cause of the decreasing discharge capacity with increasing current density is the development of a severe Li concentration gradient in the $LiV_2O_5$ film, which causes the cell to reach the cutoff voltage before the full volume of active material is utilized.

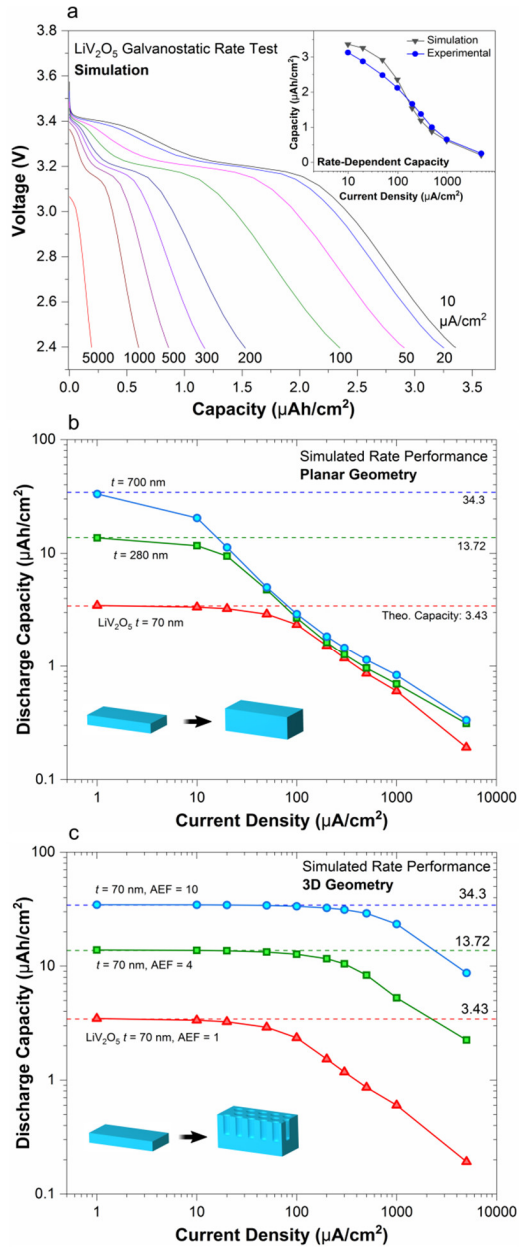

**Figure 3**: Simulation of discharge performance of $LiV_2O_5$ cathode thin films. (a) Results of optimized simulation of the galvanostatic discharge curves of a 70nm thin film $LiV_2O_5$ electrode, which can be compared to the experimental results in Figure 1f. The inset plots the predicted discharge capacity of the model vs. experiment. (b) Simulation of discharge capacity vs. current density for planar films of thickness 70nm, 280nm, and 700nm. (c) Simulation of the discharge capacity vs. current density for 3D cathodes of thickness 70nm but with area enhancement factors (AEFs) of 1, 4, and 10. All current densities are normalized to the battery footprint area.

Two ways to increase battery capacity per areal footprint are (1) increase the thickness $t$ of the capacity-limiting electrode in a planar configuration or (2) increase the internal surface area of the battery, and hence the material loading per footprint, while maintaining an optimal local electrode thickness and full self-alignment (Figure 1). The advantage of a 3D architecture in the context of footprint-limited applications can be described by the "area enhancement factor" $AEF = A/A_f$, where $A_f$ is the footprint area of the battery on the substrate and $A$ is the true total internal surface area. Trivially, a planar battery has an AEF of 1. Here we note that for the simulation results, as well as for all experimental results, reported applied current densities $i_f$ are normalized in terms of $A_f$ rather than $A$, which is the more practically-relevant metric.

Figure 3b shows the simulation results for increasing the thickness of the $LiV_2O_5$ by factors of 4 and 10. At the lowest current density (1 μA/cm²), the discharge capacity still reaches the theoretical capacity even for the 700nm thick electrode. However, performance gains from increasing the electrode thickness are rapidly lost at higher current densities, to the point that there is effectively no improvement in deliverable energy at currents above 100 μA/cm². The fundamental reason for this is that the characteristic time for a given Li flux (i.e. current density) required to reach the Li concentration corresponding to the cutoff potential at the electrode/electrolyte interface is independent of the electrode thickness, and only a thin section of a thick electrode is utilized at high currents. As a consequence, increasing the thickness of planar solid-state battery electrodes results in rapidly diminishing returns.

In Figure 3c, the same galvanostatic current range is simulated for batteries with increasing values for the AEF but with a constant $LiV_2O_5$ thickness. The theoretical capacity per device footprint is equivalent to the devices in Figure 2b. The simulation results demonstrate that 3D structuring results in a dramatic improvement in capacity retention as a function of applied current. This is due not only to the fact that the cathode thickness is locally always 70 nm, and therefore not increasingly diffusion limited as the areal material loading increases, but also because the local current density $i$ is reduced to $\frac{i_f}{AEF}$, leading to proportional reductions in the Ohmic, charge transfer, and concentration overpotentials as the AEF increases. As a result, the AEF 10 battery maintains its theoretical discharge capacity at currents up to 100 μA/cm² while the AEF 1 device is already losing capacity at 20 μA/cm². This simultaneous improvement of discharge capacity and rate performance is the hallmark of a successfully fabricated 3D battery.

**2.4 Full Cell Fabrication**

Our strategy for fabricating and testing conformal TSSBs is schematically outlined in Figure 4a-f. With the electrode/electrolyte materials compatibility already established, the remaining challenge is to develop a procedure for depositing, isolating, and testing batteries grown via conformal deposition techniques. We first fabricate 3D structures by etching hexagonal arrays of cylindrical pores into a Si substrate using deep reactive ion etching (DRIE). The array has the following dimensions: pore diameter of 3μm, center-to-center spacing of 6μm, and depth of either 12 or 30μm. The AEF of a hexagonal array of cylindrical pores with diameter $D$, center-to-center spacing $s$, and depth $h$ can be found to be $AEF = 1 + \frac{2\pi\sqrt{3}}{3}\frac{Dh}{s^2}$, leading to an expected AEF of 3.9 for the 12um pores and 9.7 for the 30um pores. For the purposes of labeling and due to some uncertainty in the exact surface area due to a scalloping effect from the DRIE, we refer to these structures as AEF 4 and AEF 10.

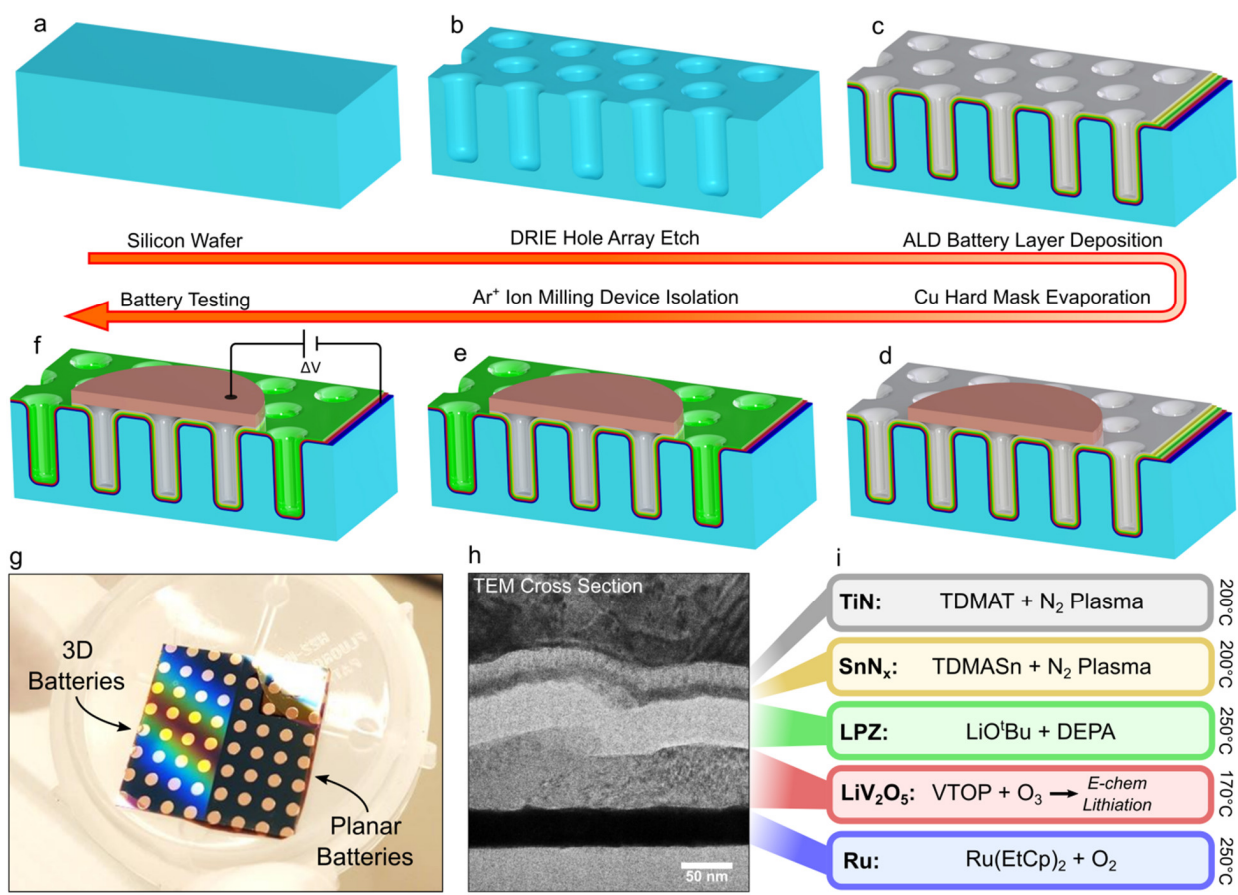

**Figure 4:** Fabrication and characterization of 3D solid state thin film batteries. (a-d) Schematic of fabrication of devices. (a) The silicon starting substrate. (b) Formation of cylindrical pore arrays via photolithographic patterning and deep reactive ion etching (DRIE) of Si. Pores are 3μm wide and either 12 (AEF 4) or 30 (AEF 10) μm in depth. (c) Blanket deposition of five active device layers via ALD, including electrochemical lithiation of the cathode as discussed in the text. (d) Deposition of Cu through a shadow mask to form 1mm diameter circular dual purpose etch mask/ needle probe contacts. (e) Isolation of individual batteries via Ar⁺ ion milling through anode current collector and anode films. (f) Battery testing through contact with top electrode and cathode current collector layers. (g) Optical photograph of finished battery "chip". Each chip is dual sided, with 3D batteries on the left and planar batteries on the right. Optical iridescence from the 3D array causes the visible coloration. (h) Cross-sectional TEM image of an all-ALD solid state battery with 40nm Ru/70nm LiV$_2$O$_5$/50nm LPZ/ 10nm SnN$_x$/ 25nm TiN. (i) Overview of ALD chemistry and process temperature for each layer visible in (h).

The high conformality and deposition temperatures associated with ALD generally prevent the use of conventional photolithography or shadow-masking. To circumvent this, we first deposit all 5 battery layers without patterning. After conformal fabrication of the battery stack, we utilize the shadow-masked PVD deposition of circular Cu etching masks with a 1mm diameter, which also serve as robust electric contacts, followed by etching of the anode and anode current collector via $Ar^+$ ion milling to isolate individual batteries. Each cell can then be tested via probe contact with an exposed area of the blanket Ru bottom-layer cathode current collector and a Cu pad. The cells are tested without further encapsulation in an Ar-filled glovebox. We fabricate dual-sided battery "chips", shown in Figure 4g, with one side containing 3D cells and the other planar cells. This allows for every tested 3D configuration to be compared 1-to-1 with planar cells made from the same deposition runs, so that any differences in performance can be reliably attributed to the cell morphology alone.

A typical all-ALD battery stack is shown in the TEM cross section in Figure 4h and the ALD chemistries used to deposit it are outlined in Figure 4i. In order of deposition, the battery is formed from 40 nm of Ru, 70nm of prelithiated $V_2O_5$ ($LiV_2O_5$), 50nm LPZ, 10nm $SnN_x$, and 25 nm TiN, finally covered in a layer of electron-beam evaporated Cu. The entire synthesis process takes place at or below 250C. The ALD LPZ is able to form a conformal and pinhole-free layer at thicknesses as low as 40nm, leading to a 100% tested device yield for planar batteries in terms of electrical isolation between anode and cathode. The achievable level of downscaling of the solid electrolyte is of interest for decreasing both cell impedance and deposition time. We previously established that approx. 30 nm LPZ was the lower limit for operation of a $LiCoO_2$/Si couple, and we observe similar trends for the $LiV_2O_5$/$SnN_x$ couple.[21] The initial yield for 3D cells depends on the exact process conditions and aspect ratio, but generally requires thicker LPZ films (>80 nm) to reach 100%. 3D cells made with very thin LPZ are much more sensitive to failure through the development of space charge limited electronic conduction (SCLC), possibly as a result of field-enhancing corners and asperities produced during DRIE.[13,19]

The battery layers are fully conformal in pore structures with an aspect ratio of 10, as indicated by SEM and energy dispersive spectroscopy (EDS) based characterization of the cross section of a cleaved AEF 10 chip (Figure 5). Figures 5a-b show SEM images of the battery stack at the top corner and bottom corner of one pore, which the locations highlighted in Figure 5c. This particular 3D cell was made using 40nm LPZ, visible as the dark layer in the film stack, and measurement of the layer thickness shows little or no change in the thickness of the LPZ from top to bottom. This 3D cell was operable (Figure S2). The visible uniformity of all 5 layers highlights the self-alignment property of ALD deposition, in contrast to previous attempts at 3D devices.[13] EDS line scans at the top, middle, and bottom of the pore, shown in Figure 5d-f, demonstrate the presence of a representative element of each of the 5 active layers throughout the pore, further supporting the conformality of the synthesis process.

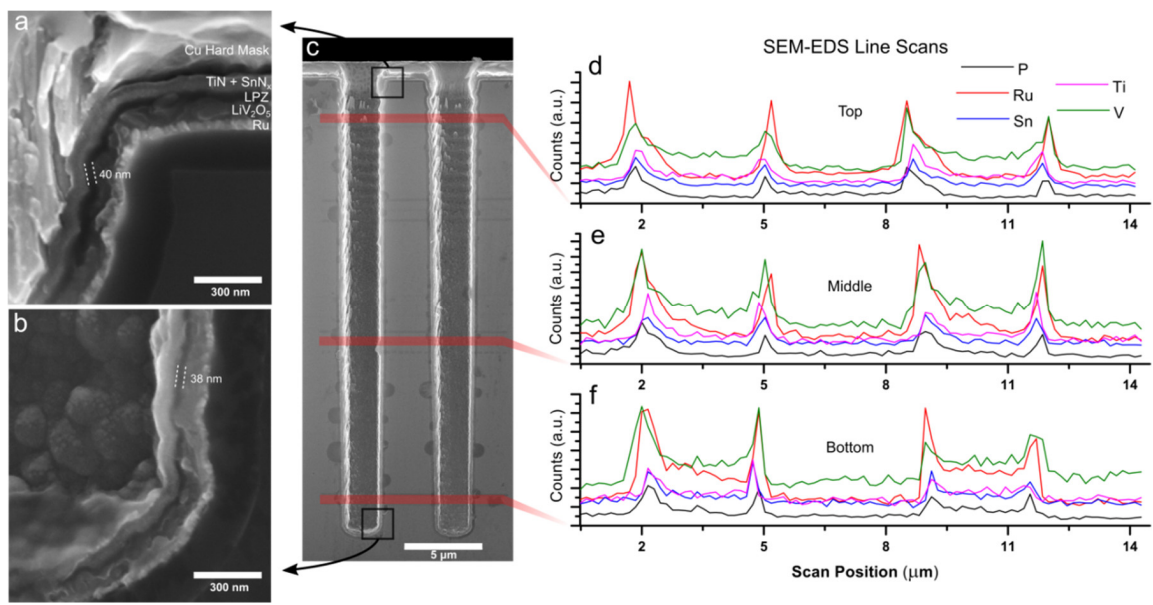

**Figure 5:** Scanning electron microscopy (SEM) and energy dispersive spectroscopy (EDS) cross-sectional characterization of a working ALD full cell (40nm Ru/70nm $LiV_2O_5$/40nm LPZ/25nm $SnN_x$/25nm TiN) grown into an AEF 10 structure. Data are taken from a battery chip cleaved along one row of holes. (a-b) SEM images of the top and bottom corners of a single cylindrical pore, shown in full length in (c). The battery layers are fully conformal down the length of the pore, including the LPZ electrolyte. (d-f) SEM-EDS line scans of the elemental concentration of P, Ru, Sn, Ti, and V from the top (d) middle (e) and bottom (f) of two pores. Peaks are associated with the increased effective sample depth at the pore walls. Each element is present throughout the depth of the pores.

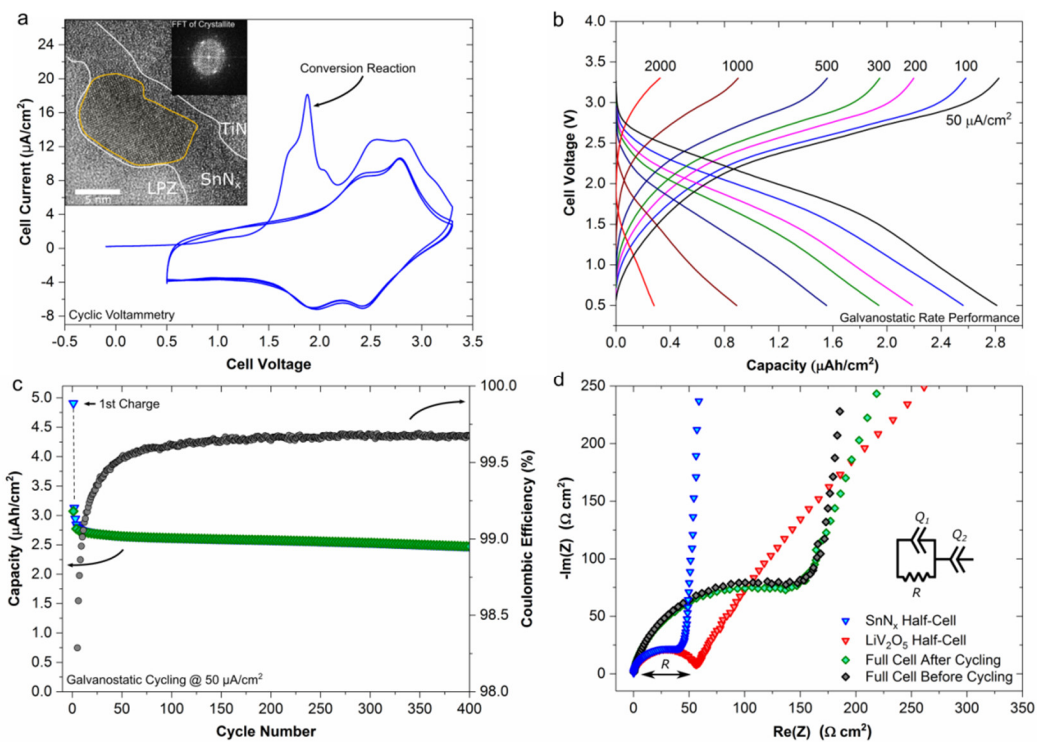

**Figure 6:** Characterization and performance of all-ALD planar solid-state cells. (a) The first three cycles of 1 mV/s cyclic voltammetry of the all-ALD chemistry between 0.5 and 3.3V. The inset shows cross-sectional TEM of a cell charged to 3.3V @ 50 uA/cm$^2$. 5-10nm crystallites, indicated by the yellow outline and corresponding fast-fourier image transform (FFT), form in the SnN$_x$ layer after the first charge, supporting a conversion-type reaction mechanism. (b) Characteristic galvanostatic charge-discharge curves using current densities between 50 and 2000 μA/cm$^2$. (c) Cycling data showing the charge capacity, discharge capacity, and Coulombic efficiency of 400 cycles at 50 μA/cm$^2$. (d) Potentiostatic electrochemical impedance spectroscopy (PEIS) of the as-made all-ALD full cell compared to the impedance of the half-cells characterized in Figure 2.

**2.4 Electrochemistry of all-ALD Solid State Cells**

Before discussing the effects of 3D structuring on performance, we first discuss the electrochemical properties of the novel LiV$_2$O$_5$ – LPZ – SnN$_x$ system itself. The cells work largely as expected based on the half-cell tests, indicating successful fabrication, though we observe an unexpected increase in cell impedance as well as an anomalous first-cycle charging capacity.

Figure 6 shows electrochemical data from approximately capacity-matched cells, with 70nm LiV$_2$O$_5$ and 10nm SnN$_x$. Figure 6a shows the first 3 cycles of cyclic voltammetry at 1 mV/s between 0.5 and 3.3V on a planar full cell. The overall characteristics, including peak locations and shapes, correspond well to a convolution of the half-cell data in Figure 2a. The prominent peak observable during the first charging sweep at 1.8V corresponds to the conversion reaction of the anode, and does not recur after the first cycle. In order to confirm this, we galvanostatically charged a cell to 3.3V before removing a lamella cross section using FIB and characterizing the SnN$_x$ layer using TEM. Shown in the inset by direct TEM imaging and a fast Fourier transform of the highlighted region, we observe the production of 5-10nm crystallites embedded in an amorphous layer, consistent with the conversion reaction producing LiSn alloys outlined earlier as well as with other TEM examinations of Sn-based conversion materials.[46] After the conversion, the full cells display the characteristic doublet of LiV$_2$O$_5$, broadened due to convolution with the Li insertion/deinsertion peaks of the anode, with peaks corresponding to the ε – α and δ – ε transitions at 2.42 and 1.97 V.

High rate galvanostatic testing between 50 and 2000 µA/cm$^2$ (Figure 6b) and cycling at 50 µA/cm$^2$ (Figure 6c) demonstrates that the full cell achieves good rate performance and is remarkably stable for 400 cycles given that it utilizes a conversion/alloying anode. The reversible capacity stabilizes after a few dozen cycles at approximately 2.6 µAh/cm$^2$, which corresponds to 37 µAh/cm$^2$·µm normalized to the cathode thickness. This value represents 75% of the theoretical capacity of the cathode as well as 53% of the theoretical capacity of the state-of-the-art sputtered LiCoO$_2$/Li couple, even after the initial conversion reaction. The Columbic efficiency stabilizes at 99.7% and the observed capacity fade is 0.02%/cyc between cycles 50-400, likely due to gradual Li loss through reactions with trace atmospheric species as the batteries are not encapsulated. We note that full-cell tests of capacity matched Li-ion cells are particularly stringent, as there is no tolerance for irreversible Li loss as there is in half cells using Li anodes in excess.

Surprisingly, the first-cycle charging capacity always significantly exceeds the theoretical capacity of LiV$_2$O$_5$. The first cycle capacity in Figure 6c is 49 µAh/cm$^2$·µm, a 42% excess over the initial capacity of the cathode. This excess capacity is fortuitous as it does not appear to impede the operation of the full cell, and to some degree compensates for the first-cycle conversion reaction losses. We first considered whether the LPZ deposition chemistry was providing excess Li to the LiV$_2$O$_5$ cathode through direct lithiation by LiO$^t$Bu, as has been observed.[30] However, only an 11% excess was observed for the half-cell (Figure 2f), which should show a similar effect, and V 2p core level XPS taken of the LPZ/LiV$_2$O$_5$ interface after deposition of a few nm of LPZ shows no additional reduction in V valence state, which would be associated with extra Li insertion (Figure S3). While previous testing showed ALD LPZ to be electrochemically stable between 0.1 and 3.8V vs. Li/Li+ on a Pt electrode, we propose that the SnN$_x$ conversion reaction partially consumes adjacent electrolyte, which supplies the excess Li. The smaller excess in the half-cell could arise from different decomposition reactions with lithium metal.

Potentiostatic impedance spectroscopy (PEIS) reveals that the full cell exhibits an internal impedance anomalously higher than expected based on half-cell testing. Figure 6d shows Nyquist plots of the impedance of a full planar cell (with 40 nm LPZ) before and after cycling, as well as the impedances of the half cells tested in Figure 2 (with 80 nm LPZ). The semicircle at high frequencies (lower left of graph) reflects the ionic conductivity of the electrolyte, which is determined via fitting the data with model shown. The model includes the electrolyte resistance $R$ in parallel with a constant phase element Q, with an additional constant phase element to model the blocking response at lower frequencies. The first-cycle conversion reaction as well as cycling of the full cells does not increase the cell resistance relative to their initial state, supporting the high reversibility of the couple. However, the fit value of $R$ in the full cell (172 Ω cm$^2$) is significantly higher than that of the half cells (56 and 47 Ω cm$^2$ for the cathode and anode cells respectively) despite using an electrolyte half as thick. The only differences in fabrication and processing for the full cells are the growth of the SnN$_x$ directly on the LPZ, as well as the use of ion milling. As ALD processes are known to sometimes induce substrate damage,[47] we examined both the LiV$_2$O$_5$/LPZ and LPZ/SnN$_x$ interfaces directly by growing a very thin overlayer and directly characterizing the interface chemistry with XPS (Figure S3 and S4). Both interfaces seem well-preserved, with only some alterations of the N chemistry within the LPZ observed at both interfaces. The origin of the additional impedance requires further investigation.

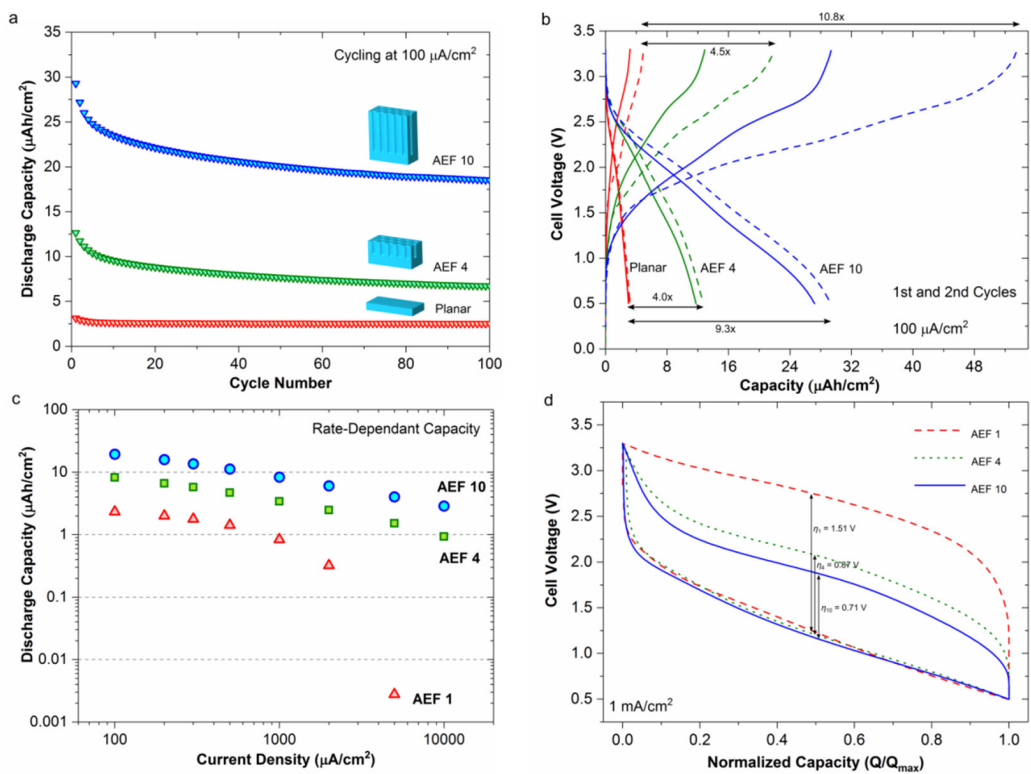

**Figure 7:** Electrochemical performance of 3D solid state batteries. (a) Cycling performance of AEF 1, 4, and 10 batteries galvanostatically cycled 100 times at 100 µA/cm2. (b) First and second charge and discharge profiles of AEF 1, 4, and 10 batteries. The arrows show the measured capacity enhancement of the AEF 4 and 10 devices relative to the AEF 1 (planar) battery, with the upper arrows showing the enhancement factors measured for the first charge and the lower arrows indicating those for the first discharge. (c) Discharge capacity as a function of the applied current density for AEF 1, 4, and 10 batteries. Data were taken after a burn-in process, i.e. after the majority of the rapid capacity loss observable in the first 50 cycles in (a). (d) Cell voltage vs. normalized capacity ($Q/Q_{max}$) for AEF 1, 4 and 10 batteries cycled at 1 mA/cm$^2$ after burn-in. The arrows indicate the measured overpotential η at $Q/Q_{max}$ = 0.5.

### 2.5 3D Full Cell TSSBs

Having successfully established a viable solid state battery from an electrochemistry and process chemistry standpoint, we turn to the concept for which conformal deposition is a unique enabler- 3D architectures. We successfully integrated the full cell into 3D substrates with AEF 4 and AEF 10, and found the footprint-normalized battery performance to be dramatically improved in terms of capacity, rate performance, and round-trip efficiency (RTE). These batteries represent the first example of operating, self-aligned solid-state batteries grown by conformal (chemical) vapor phase deposition of any kind, and serve as a benchmark for future optimization. Figure 7a shows 100 galvanostatic cycles at 100 µA/cm$^2$ between 3.3 and 0.5V for a planar, an AEF 4, and an AEF 10 cell, with Figure 7b displaying the first and second charge/discharge curves from the same data. These cells were constructed with the standard 70nm $LiV_2O_5$/10nm $SnN_x$ loading using 100nm LPZ as the solid electrolyte (Figure S5). During the initial cycles, the device performance meets the theoretical geometric enhancement. The measured capacity enhancement, shown by the horizontal arrows, of the first charge relative to the planar reference cell is 4.5x for the AEF 4 battery and is 10.8x for the AEF 10 battery, followed by 4x and 9.3x, respectively, for the first discharge. This is direct evidence of (1) the uniformity of the battery layers within the 3D geometry as well as (2) the ability of the solid-state electrolyte to provide full electrical isolation in 3D structures.

However, we consistently observe a more rapid decay in capacity, especially for the first ~10 cycles, for 3D cells vs. planar cells. By the 100$^{th}$ cycle, the discharge capacity enhancement has declined to 2.6x for the AEF 4 cell and 7.3x for the AEF 10 cell. We note that there is one additional difference in device architecture other than the increased surface area in 3D cells- the Cu capping layer is no longer covering the full active area of the battery, as it is not conformal. Inside the pores, the topmost layer is primarily the TiN current collector (Figure 5b). If the TiN layer is not acting as a perfect Li diffusion barrier, free Li may be diffusing to the surface and irreversibly forming reaction products with atmospheric reactants. This hypothesis is supported by the fact that the average discharge potential of the 3D full cells also declines with cycle number (Figure S6). A loss of active Li entirely out of the cell (rather than remaining irreversibly trapped inside the anode) would lead to a decline in cell voltage as a result of the average anode potential sliding upwards along the CV curves shown in Figure 2a. Direct Li entrapment in the TiN itself is unlikely, as this would also lead to significant capacity fade in the planar cells. Capacity loss due to a lack of high quality encapsulation in thin film SSBs is a well-known problem; future development must include better packaging for 3D geometries.[4,11]

The ultimate test of a 3D architecture is the ability to maintain deliverable capacity with applied current densities beyond the reach of planar architectures. Figure 6c plots the rate performance of the three tested geometries between 0.1 and 10 mA/cm$^2$. In this current range, simulation predicts that improving cell performance through increasing the cathode thickness in a planar configuration is impossible (Figure 2c). In order to prevent convolution with early-cycle capacity fade as the rate was varied, these data were taken after a "burn in" process of multiple slow CV cycles in order to stabilize the capacity of the 3D cells. As can be seen, 3D structuring results in better capacity retention at higher current densities while simultaneously improving total discharge capacity. The planar cells are immediately polarized to the cutoff potential at currents above 2 mA/cm$^2$, whereas the AEF 10 cell is still able to deliver a discharge capacity greater than the full capacity of the planar cells even when cycled at the exceptionally high current density of 10 mA/cm$^2$. A comparison of Figure 7c and Figure 3b-c clearly demonstrates the experimental 3D cells are operating qualitatively, if not quantitatively, within the favorable scaling regime identified by simulation, with the primary deviation arising from the fact that the simulation assumes a Li anode and does not account for first-cycle irreversibility or subsequent capacity fade.

3D structuring also significantly improves the round-trip efficiency (RTE) of batteries cycled at moderate to high rates through the reduction of the internal current density, which reduces both Ohmic and charge transfer overpotentials. Conversion-type electrode materials commonly suffer from a low RTE.[48] Plotting a full cycle at 1 mA/cm$^2$ for each of AEF 1, 4, and 10 batteries with the capacity $Q$ normalized to the achieved capacity at the cutoff potential of 0.5 V $Q_{max}$ (Figure 6d) reveals the progressive reduction in hysteresis. At the halfway point $Q/Q_{max} = 0.5$, the charge-discharge hysteresis overpotential $\eta$ is reduced from 1.51V for AEF 1 to 0.71 V for AEF 10, and the net RTE is improved from 45% to 64%.

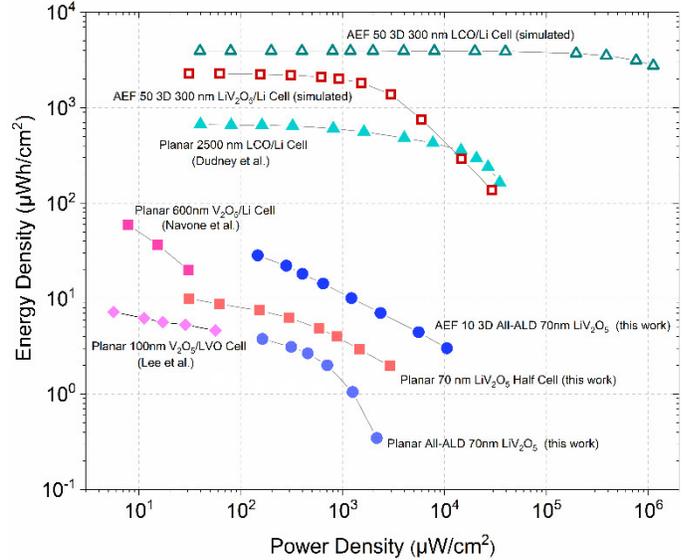

**Figure 8:** Ragone plot of device performance for various TSSB configurations. Squares denote data from LiV$_2$O$_5$/Li cells, circles from LiV$_2$O$_5$/SnN$_x$ cells, and triangles from LiCoO$_2$/Li cells. Solid symbols denote experimental data while outlined symbols (the topmost two curves) denote data from COMSOL simulations as described in the supplementary information. Data for the 600nm V$_2$O$_5$/Li cell is extracted from Ref. 49, data for the 100nm V$_2$O$_5$/LVO cell was estimated from ref. 50, and data for the 2500nm LCO/Li cell is extracted from Ref. 4.

## 3. Prospects for Architecture Scaling

The planar and 3D batteries described in this work are promising from an electrochemistry standpoint, establish a path towards a new performance regime for solid state storage, and represent the highest powers tested for vanadium oxide-based solid-state cells. Here we briefly place the experimental results into context and explore future opportunities for performance improvement. Areal energy density vs. average areal power density (derived from galvanostatic tests in all cases) for a variety of architectures both experimental and simulated is plotted in Figure 8. The AEF 10 3D all-ALD battery is superior to both the planar LiV$_2$O$_5$ and all-ALD cells, as expected, as well as to literature references for vanadium oxide-based TSSBs. Data derived from a 600nm V$_2$O$_5$/LiPON/Li TSSB described by Navone et al. demonstrate[49] clearly the benefit of 3D structuring; a rough extrapolation of the rate performance of the cell shows that it would, at best, perform similarly to the planar all-ALD device at higher power densities despite the ~8.5x thicker cathode. The planar all-ALD cell also performs at least as well as the best-characterized example of a $V_2O_5$ lithium-ion TSSB, which used a deeply lithiated vanadium oxide (LVO) film as the anode.[50]

While this work represents a significant step forward in terms of 3D battery fabrication, we also wish to establish a path towards exceeding the best existing TSSBs in terms of absolute areal performance metrics. The absolute performance of these proof-of-concept designs cannot yet compete with the best examples of RF-sputtered planar SSBs using thick, highly crystalline $LiCoO_2$ cathodes due to the low diffusivity of Li ($D^* \approx 3 \times 10^{-13} cm^2/s$) in ALD-grown $LiV_2O_5$. The average $D^*$ for $LiCoO_2$ can reach over $10^{-10} cm^2/s$, leading to extraordinary power performance,[51,52] although reaching this value requires high temperature annealing which can impede device integration or substrate compatibility. Experimental data from a 2500nm $LiCoO_2$/LiPON/Li cell developed by Dudney et al. (Figure 8), which operates with over 50% capacity retention at power densities of over 10 mW/cm$^2$, provides a benchmark.[4]

Straightforward methods of optimizing the 3D cells include increasing the AEF through etching higher aspect ratio structures and packing them more closely, as well as increasing the thickness of the cathode film and/or replacing the cathode with a different material entirely. Replacing the anode with Li or Si, which may be possible in specific lower AEF configurations with CVD or simple melt-impregnation, would also improve the cell voltage and reduce the first-cycle irreversibility. We include in Figure 8 simulations of two architectures which we argue represent reasonable upper bounds for ALD-grown 3D microbatteries. In principle, there is no limit to either the thickness of films grown by ALD or the aspect ratio in which they can be deposited. However, growing films more than a few hundred nm in thickness by ALD is likely impractical due to the slow rate of deposition, and because the precursor dose required for saturated growth scales as approximately the square of the aspect ratio,[53] batteries with an AEF of more than ~100 would be extremely challenging to fabricate. AEFs of ~50 for an ALD $TiO_2$ half-cell have been recently demonstrated, so we use this value as an achievable goal.[23] The simulations assume the use of a Li anode for simplicity; the use of $SnN_x$ as the anode would reduce the energy density by approximately ~2x based on the experimental results.

The simulation results (Figure 7, topmost curves) indicate that an AEF 50 battery using a 300nm $LiV_2O_5$ cathode and a 100nm LPZ solid electrolyte would significantly exceed the energy density of existing LCO-based TSSBs and reach the mWh/cm$^2$ range required to compete with existing Li-ion conventional batteries. However, $LiV_2O_5$-based cells are likely not capable of besting thick LCO-based planar TSSBs at power densities above 10 mW/cm$^2$. Even at AEF 50, the difference in $D^*$ is too great. Truly mold-breaking batteries require replacing the $LiV_2O_5$ cathode with conformally-grown high quality LCO. Simulations of an AEF 50 battery with 300nm LCO, assuming a typical $D^*$ of $10^{-10}$ cm$^2$/s, yield a solid-state device on the verge of competing with conventional Li-ion cells in terms of energy density (3.9 mWh/cm$^2$) and which can maintain 90% energy retention at a power density of 386 mW/cm$^2$ (corresponding to an approximate C-rate of 110). Conformal deposition of high quality LCO may be possible through further optimization of an ALD process[31] or through electrodeposition,[54] and will be explored as a next step. In addition, some reports of well-crystallized $V_2O_5$ electrodes measure values for $D^*$ above $10^{-11}$ cm$^2$/s (likely dependent on crystalline orientation), and so it may be possible to increase the performance of ALD-grown $LiV_2O_5$ with additional treatments or process modification.[55] The energy density of the all-ALD cells can also be significantly increased by more deeply prelithiating the ALD $V_2O_5$ to $Li_2V_2O_5$, which preliminary experiments indicate is also a promising approach.

## 4. Conclusions

3D structuring of thin film solid state batteries is a promising method of producing high-performance, intrinsically safe energy storage devices with exceptional areal energy and power densities. For the first time, we have established a set of materials (a prelithiated $LiV_2O_5$ cathode, a $SnN_x$ anode, and a lithium polyphosphazene solid electrolyte) which are mutually compatible, are grown in the active phase at temperatures ≤ 250C, and can be reliably made using conventional ALD deposition tools which are now common in industrial and university settings. 3D cells can be successfully fabricated through deposition in micromachined silicon substrates followed by masked etching, and full electrical isolation between anode and cathode can be achieved with solid electrolytes ≤ 100nm in thickness in structures with an AEF of up to 10. Solid state batteries made from the $LiV_2O_5$-$SnN_x$ couple exhibit stable capacities of 2.6 µAh/cm$^2$, (37 µAh/cm$^2$·µm normalized to the cathode thickness) for hundreds of cycles. The areal discharge capacity of these cells can be scaled up to 9.3x that of planar cells through integration with 3D substrates, though at present 3D cells suffer from additional anomalous capacity loss that should be addressed through better cell encapsulation. Most importantly, 3D structuring improved the rate performance and RTE of the cells while simultaneously increasing the areal capacity. This beneficial combination was measured in a range of current densities ($i_f \geq 100$ µA/cm$^2$) which was indicated by simulation to be a power regime in which such scaling was not possible for planar cells.

Future development of 3D TSSBs can utilize a "mix-and-match" strategy for materials selection combined with the fabrication scheme developed in this work, though at the present time the ALD LPZ electrolyte is probably the best conformal inorganic electrolyte available. While the use of LiV$_2$O$_5$ may continue to be appropriate for integration with temperature-sensitive substrates such as polymer films, matching and exceeding the performance of conventional Li-ion cells will likely require its replacement with a cathode material with a higher chemical diffusion constant for Li, such as LiCoO$_2$. It would also be interesting to explore the integration of the conformal TSSBs described here with more extensively three-dimensional substrates, such as fabrics, fibers, conductive metal foams, which could form the basis of multifunctional energy-storing materials and composites.

## Author Contributions

A.J.P, G.W.R., and K.E.G. conceived and designed the research program. A.J.P. fabricated and characterized devices and developed the computational models. T.E.S., E.S., and D.S. contributed to materials development and characterization. A.C.K. and K.G. contributed to research infrastructure. A.J.P and K.E.G. wrote the manuscript.

## Acknowledgements

All aspects of this work were initiated and supported by Nanostructures for Electrical Energy Storage (NEES), an Energy Frontier Research Center (EFRC) funded by the U.S. Department of Energy, Office of Science, Office of Basic Energy Sciences under award number DESC0001160. This work builds on regular discussions within the NEES T4 Solid State Storage thrust, and frequent extensive discussions with Henry White and Kim McKelvey (U. Utah) as well as Talin and Bruce Dunn (UCLA). Some aspects of the 3D array microfabrication process were partially supported by Independent Research and Development Funding from the Research & Exploratory Development Department (REDD) of the Johns Hopkins University Applied Physics Laboratory (JHU/APL). We appreciate the facilities and support within the Maryland NanoCenter, including its Fablab for device fabrication and its AIMLab for microscopy and FIB. Ion milling was performed at the NIST Center for Nanoscale Science and Technology.

# Supplementary Section 1: Experimental Methods

**Device Fabrication**

All samples were fabricated using Si test wafers as a starting material. The device footprint of all tested electrochemical devices (half cells, full cells, and 3D cells) was defined by a 1mm diameter circular contact pad. Planar half-cell devices were constructed from diced Si wafers coated with a 70nm Pt current collector deposited via electron-beam deposition with a 5nm Ti adhesion layer. For half cells 3μm thick Li metal electrodes were deposited using thermal evaporation through a stainless steel shadow mask in a homebuilt vacuum evaporator coupled to an Ar filled glovebox. 3D substrates were fabricated via the formation of etch masks via standard photolithographic patterning followed by deep reactive ion etching (DRIE) using a Bosch process in an STS etching system. Etched wafers were RCA-cleaned and subsequently thermally oxidized in a Tystar CVD system to form a 200nm $SiO_2$ layer, serving as a Li diffusion barrier and pristine surface for ALD growth. After ALD deposition/electrochemical formation of the 5 active layers, individual batteries were defined by depositing via electron beam deposition 1μm of Cu through a shadow mask, acting as a probe contact and etch mask. After Cu deposition, low energy $Ar^+$ ion milling with SIMS-based endpoint detection (4Wave Systems) was used to etch the TiN and $SnN_x$ layers, which electrically isolated each top contact. Samples are briefly air-exposed after the formation of the cathode layer, but further synthesis and characterization is performed entirely in vacuum or Ar environments.

**Active Layer Formation**

Ruthenium metal was grown in a homebuilt tube-furnace type reactor using $Ru(EtCp)_2$ and $O_2$ at 250˚C. Crystalline $V_2O_5$ was grown in a Beneq TFS 500 ALD reactor using vanadium triisopropoxide (VTOP) and $O_3$ at 170˚C using an optimized variant of a previously described process.[1] After $V_2O_5$ deposition, $LiV_2O_5$ was formed via galvanostatic electrochemical insertion of Li in a 0.25M $LiClO_4$/propylene carbonate (PC) electrolyte with a Li metal counter electrode at a C/3 rate, with a cutoff of 2.8V vs. Li. Excess electrolyte/salt was removed by briefly soaking the sample in pure PC and rinsing with isopropanol. The lithium polyphosphazene (LPZ) solid electrolyte was grown at 0.6Å/cyc in a Fiji F200 ALD reactor using lithium *tert*-butoxide and diethyl phosphoramidate as reactants at 250˚C.[2] The LPZ thickness ranged from 40 – 100 nm for various devices. For 3D substrates, an exposure process was used in which a butterfly valve shut off active pumping to the chamber during precursor pulses to ensure full conformality. Following LPZ deposition, the samples were transferred without air exposure into a second Fiji F200 ALD reactor. The $SnN_x$ anode was deposited at 200˚C using tetrakis(dimethylamido)tin (TDMASn) and a $N_2$ plasma with a growth rate of 0.5Å/cyc, followed by deposition of TiN using tetrakis(dimethylamido)titanium (TDMAT) and a $N_2$ plasma, also at 200˚C and with a similar growth rate. Layer thicknesses were measured by SEM cross section and have an estimated error of 10%.

**Characterization**

X-ray photoelectron spectroscopy (XPS) was performed using a Kratos Axis Ultra DLD spectrometer vacuum coupled to both the ALD reactor and glovebox used for device synthesis and testing, which allowed for characterization without air exposure. XPS was taken using a monochromated Al x-ray gun operated at 144W using pass energies between 160 and 20 eV. Spectra were calibrated, when possible, to a hydrocarbon C 1s peak at 284.8 eV. XPS peak fitting was done using CasaXPS using Shirley or linear backgrounds and 50/50 Voight-type lineshapes using appropriate area ratios for spin-orbit split components. Focused ion beam processing and scanning electron microscopy were performed using a Tescan XEIA FEG SEM dual beam system. Transmission electron microscopy was performed using a JEM 2100 FEG TEM. Individual batteries were tested in an Ar-filled glovebox via connection with custom-built micromanipulator probe contacts with coaxial connections to an external Biologic VSP potentiostat with an impedance channel. PEIS data were taken at room temperature with an excitation amplitude of 50mV.

# Figure S1: Comparison of ALD SnO$_2$ and ALD SnN$_x$ Half-Cells

We tested solid state half cells of both SnN$_x$ and SnO$_2$ thin films (grown via reaction between TDMASn and H$_2$O) and found the nitride to be the superior thin film anode. Figure S1 shows the first cycle of cyclic voltammetry of solid state half cells made with ALD tin nitride and oxide. The oxide film delithiates in a complex process exhibiting multiple peaks, including a substantial fraction of the capacity above 1.2 V vs. Li/Li+. In contrast, SnN$_x$ delithiates in a single process at approx. 0.9V vs. Li/Li+ leading to an increased full cell operating voltage. The nitride also exhibits superior cycling stability, described fully in a separate publication.[3]

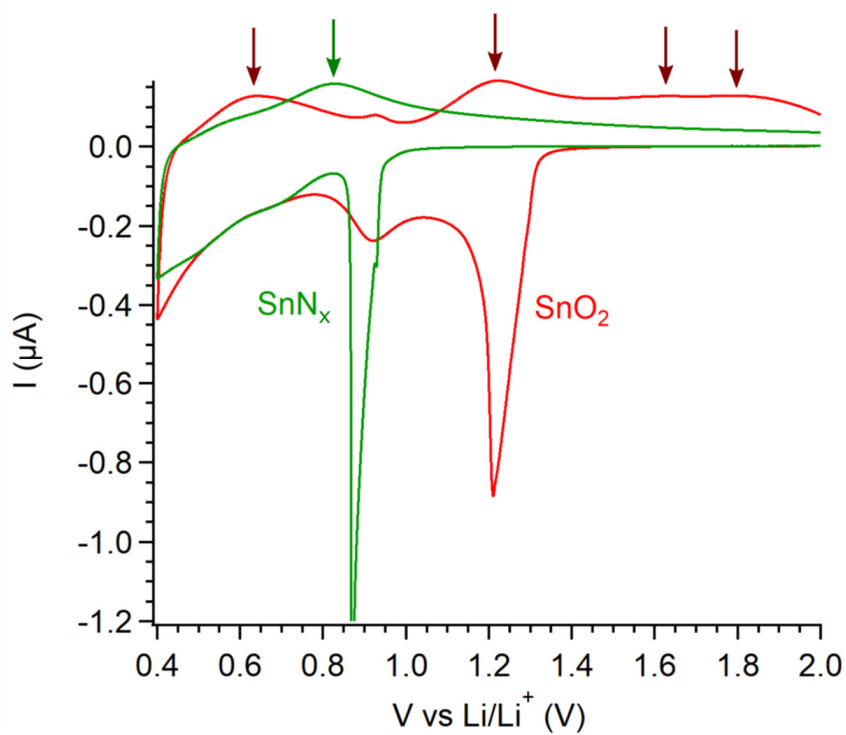

**Figure S2:** Typical cyclic voltammetry of ALD-grown SnN$_x$ and SnO$_2$ thin films paired with an ALD LPZ solid electrolyte and a Li metal anode. The scan rate is 2 mV/s.

# Supplementary Discussion 1: Details of Finite Element Simulation

A 1D time-dependent finite element model of several types of thin film lithium ion batteries was developed for use in predicting performance trends of 3D architectures. The models were implemented and solved using COMSOL Multiphysics 5.2a. Two slightly different models were developed for batteries using a LiV$_2$O$_5$ or LiCoO$_2$ cathodes based on a blend of empirical and literature data. The models erred on the side of simplicity unless high-quality empirical or previously computationally optimized parameters were available. The model has three primary elements: (1) Li transport in the electrolyte (2) charge transfer at the electrode/electrolyte interface and (3) Li transport in the electrode.

**Transport in the electrolyte**: Transport physics were based on the equations proposed and compared with experimental data for LiPON-based TSSBs in papers by Danilov et al.[4,5] The movement of charged species is governed by the Nernst-Planck equation with electroneutrality:

$$\frac{\partial c_i}{\partial \tau} = \nabla \left( -D_i^* \nabla c_i + \frac{F z_i}{RT} D_i^* c_i \nabla \varphi_l \right)$$

$$\sum_i z_i c_i = 0$$

where $c_i$ is the concentration, $D_i^*$ is the chemical diffusion coefficient, and $z_i$ is the electric charge of the $i^{\text{th}}$ species. $F$ is Faraday's constant, $R$ is the gas constant, $\tau$ is time, $\varphi_l$ is the electric potential in the electrolyte, and $T$ is the temperature.

The solid electrolyte is assumed to contain a fixed concentration of Li atoms $c_0$, of which a fixed fraction $\varepsilon < 1$ are ionized, mobile charge carriers in equilibrium through the reaction

$$\text{Li}^0 \leftrightarrow \text{Li}^+ + \text{a}^-$$

where a$^-$ is a compensating negatively charged species[1] or defect, and which has forward and backward rate constants of $k_d$ and $k_r$, respectively. The net dissociation rate $G$ is then

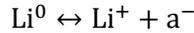

$$G = k_d c_{\text{Li}^0} - k_r c_{\text{Li}^+} c_{\text{a}^-}$$

which, when combined with the electroneutrality condition, leads to the relation

$$k_d = \frac{k_r c_0 \varepsilon^2}{(1-\varepsilon)}$$

We do not measure $D_{\text{Li}^+}^*$ directly, but instead calculate it from experimental measurements of the ALD LPZ ionic conductivity $\sigma$ and the molar concentration $c_0$ of Li ions, calculated from the stoichiometry measured via XPS and density measured via x-ray reflectometry. The Nernst-Einstein relation links these quantities:

---

[1] It may be noticed that the model assumes the species a$^-$ to have a value for $D_{\text{a}^-}^*$ comparable to that for $D_{\text{Li}^+}^*$, which is physically somewhat surprising for a solid electrolyte. Despite previous studies finding good agreement between this model and experimental data using similar values for $D_a^*$, it is not clear what species would meet this condition, considering that LiPON-family solid electrolytes are excellent electronic insulators and do not contain obvious candidates for mobile negatively charged ions. Altering the ratio $D_{\text{a}^-}^*/D_{\text{Li}^+}^*$ has relatively little impact on the simulated discharge curves within the experimental current density regime, but has an enormous impact on the predictions of concentration gradients within the solid electrolyte. Unfortunately, methods of measuring such distributions *in operando* in thin film devices remain elusive.

$$D^*_{Li^+} = \frac{kT\sigma}{c_0 \varepsilon e^2}$$

where $k$ is the Boltzmann constant and $e$ is the elementary charge.

**Charge transfer at electrode/electrolyte interfaces:** For a given local cell current density $i_{cell}$, coupling of the Li ion fluxes at the boundary between the cathode and electrolyte is governed by Butler-Volmer kinetics as a function of the overpotential $\eta$:

$$i_{cell} = i_0^c \left( e^{\alpha F \eta / RT} + e^{-(1-\alpha)F\eta/RT} \right)$$

$$\eta = \varphi_s - \varphi_l - E_{eq}$$

$$i_0^c = F k_c \left( (c^M_{Li^+} - c_{Li^+}) c_{Li} \right)^\alpha \left( (c^M_{Li} - c_{Li}) c_{Li^+} \right)^{1-\alpha}$$

where $i_0^c$ is the cathode exchange current density, $c_{Li}$ is the concentration of Li in the cathode, $c_{Li^+}$ is the concentration of charge carriers in the electrolyte, $\alpha$ is the charge transfer coefficient, $\varphi_s$ and $\varphi_l$ are the electric potentials in the electrolyte and electrode, respectively, $c^M_{Li}$ is the maximum Li concentration in the cathode, $c^M_{Li^+}$ is the maximum Li concentration in the electrolyte, and $k_c$ is the cathode reaction rate constant.

Despite the kinetic theory prediction that $i_0^c$ depends on $c_{Li}$, and therefore the state-of-charge, most experimental attempts to measure the charge transfer resistance find it to be relatively invariant (or, at most, following a weak trend incompatible with the function above).[6–8] For this reason, we make the approximation that

$$i_0^c = const.$$

The anode/electrolyte interface is treated similarly, with the additional condition that $E_{eq} = 0$ and $\varphi_s = 0$ to simulate a conductive, highly reversible Li metal anode.

$E_{eq}(c_{Li})$ is an empirically measured function of $c_{Li}$ and reflects the chemical potential of Li ions in the electrode at different states of charge during a quasistatic discharge. Shown below are the curves used for $LiCoO_2$ and $LiV_2O_5$:

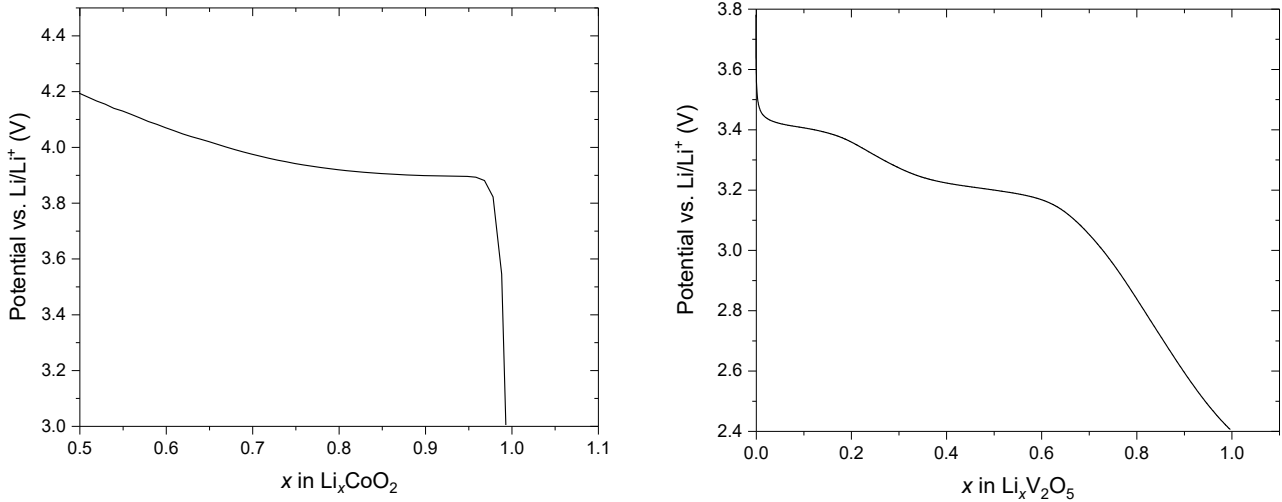

**Transport in the electrodes:** Movement of Li ions in the cathode films is modelled in one dimension using Fick's law:

$$\frac{dc_{Li}}{d\tau} = \frac{d}{dx}\left[D_{Li}^*(x, c_{Li})\frac{dc_{Li}}{dx}\right]$$

where in principle $D_{Li}^*$ can depending on position and concentration. In general, $D_{Li}^*$ is not constant as the composition of a battery material changes, but good agreement with experimental data can nonetheless be attained with constant average values. Electronic transport is neglected due to the relatively high electronic conductivity of both $LiCoO_2$ and $LiV_2O_5$.

For $LiV_2O_5$, we were unable to achieve satisfactory agreement between the model and experimental discharge curves using a constant $D_{Li}^*$ because of consistent over-performance at high current densities (i.e. the model would predict a lower-than-measured capacity). Attempts to include concentration dependence were not successful. Instead, we achieved best agreement by including a small spatial dependence to $D_{Li}^*$ where the 10% of the film adjacent to the electrode/electrode interface has a higher diffusivity:

$$\widetilde{D}_{Li,LVO}^*(x) = \begin{cases} 10 \cdot D_{Li,LVO}^*, & 0 \leq x < t/10 \\ D_{Li,LVO}^*, & t/10 \leq x \leq t \end{cases}$$

where $x$ is the position in the electrode with $x = 0$ representing the $LiV_2O_5$/LPZ interface and $x = t$ representing the external boundary of the electrode. There are two reasonable physical interpretations of this effect. The first is that such a function is an *ad-hoc* method of modelling pseudocapacitance, i.e. fast faradaic charge transfer restricted to near the electrode/electrolyte interface, which has been characterized in $V_2O_5$ previously[9] and is supported by the measured presence of some non-diffusion limited charge storage in the $LiV_2O_5$ half cell (Figure 2b in the main text). The second possibility is that this represents a 1D approximation of 2D crystalline heterogeneity in the $LiV_2O_5$ film itself. For instance, if ~10% of the $LiV_2O_5$ crystal grains had a more favorable orientation (i.e. planes with a higher Li diffusivity for Li in direct contact with the LPZ), we would observe an overperformance at high current densities, as only those properly oriented grains would be active. [8]

The value of $D_{Li}^*$ for LCO is taken as a simple constant estimated from typical literature values.[10]

**Simulating the effect of 3D structuring**: In order to model 3D architectures with a given $AEF$ using a 1D model, we assume that the current density within the 3D cells is fully homogeneous across the entire surface area of the battery due to the relatively high electronic conductivity of the conformal TiN anode current collector. In this case, the local current density $i_{cell}$ relative to the applied footprint current density $i_f$ is simply

$$i_{cell} = \frac{i_f}{AEF}$$

The expected areal capacity $Q_f$ of the modelled 3D cell is then found by multiplying the output capacity of the 1D model $Q_{cell}$ at the cutoff potential by the $AEF$:

$$Q_f = AEF \cdot Q_{cell}$$

The stop condition of the simulation is $\varphi_s(x=0) \leq V_c$, where $V_c$ is the cell cutoff voltage.

# Table S1: Model Variables and Parameters

| Quantity | Dimension | Value | Description | Source |
|---|---|---|---|---|
| $t_A$ | m | $100 \cdot 10^{-9}$ | Anode thickness | Exp. |
| $t_E$ | m | $100 \cdot 10^{-9}$ | Electrolyte thickness | Exp. |
| $t$ | m | - | Cathode thickness | Exp. |
| $\tau$ | s | - | Time | - |
| $F$ | C mol$^{-1}$ | 96485 | Faraday's constant | - |
| $R$ | J mol$^{-1}$ K$^{-1}$ | 8.314 | Gas constant | - |
| $T$ | K | 298 | Temperature | Exp. |
| $c_{Li^0}$ | mol m$^{-3}$ | - | Concentration of neutral Li atoms in LPZ | - |
| $c_{Li^+}$ | mol m$^{-3}$ | - | Concentration of mobile Li ions in LPZ | - |
| $c_{a^-}$ | mol m$^{-3}$ | - | Concentration of counter charges in LPZ | - |
| $k_d$ | s$^{-1}$ | $1.49 \cdot 10^{-5}$ | Dissociation rate constant in LPZ | Calculated |
| $k_r$ | m$^3$ mol$^{-1}$ s$^{-1}$ | $9 \cdot 10^{-7}$ | Recombination rate constant in LPZ | Ref. 4 |
| $c_0$ | mol m$^{-3}$ | $3.32 \cdot 10^4$ | Concentration of Li atoms in LPZ | Calculated |
| $\varepsilon$ | - | 0.2 | Fraction of total Li ions mobile in eq. in LPZ | Ref. 4 |
| $e$ | C | $1.6 \cdot 10^{-19}$ | Elementary charge | - |
| $\sigma$ | S cm$^{-1}$ | $6.6 \cdot 10^{-7}$ | Ionic conductivity of LPZ | Exp. |
| $c_{Li}$ | mol m$^{-3}$ | - | Concentration of Li ions in electrode | - |
| $D^*_{Li^+}$ | cm$^2$ s$^{-1}$ | $2.74 \cdot 10^{-11}$ | Chem. diffusion coeff. of Li ions in LPZ | Calculated |
| $D^*_{a^-}$ | cm$^2$ s$^{-1}$ | $5.1 \cdot 10^{-11}$ | Chem. diffusion coeff. of counter charges in LPZ | Ref. 4 |
| $D^*_{Li,LVO}$ | cm$^2$ s$^{-1}$ | $3 \cdot 10^{-13}$ | Chem. diffusion coeff. of Li ions in LiV$_2$O$_5$ | Optimized |
| $D^*_{Li,LCO}$ | cm$^2$ s$^{-1}$ | $1 \cdot 10^{-10}$ | Chem. diffusion coeff. of Li ions in LCO | Ref. 10 |
| $c^M_{Li,LVO}$ | mol m$^{-3}$ | 18431 | Max. conc. of Li in LiV$_2$O$_5$ | Estimated |
| $c^M_{Li,LCO}$ | mol m$^{-3}$ | 50000 | Max. conc. of Li in LCO | Estimated |
| $\alpha$ | - | 0.5 | Charge transfer coefficient for LCO and LVO | Estimated |
| $i_0^{LCO}$ | A cm$^{-2}$ | $4.89 \cdot 10^{-4}$ | LCO exchange current density | Ref. 6 |
| $i_0^{LVO}$ | A cm$^{-2}$ | $1.3 \cdot 10^{-4}$ | LVO exchange current density | Ref. 7 |
| $\varphi_l$ | V | - | Electric potential in electrolyte | - |
| $\varphi_s$ | V | - | Electric potential in electrode | - |
| $\eta$ | V | - | Electrode overpotential | - |
| $x$ | m | - | Position | - |
| $i_{cell}$ | A cm$^{-2}$ | - | Applied local current density | - |

# Figure S2: Cycling of a 3D Cell with 40nm LPZ

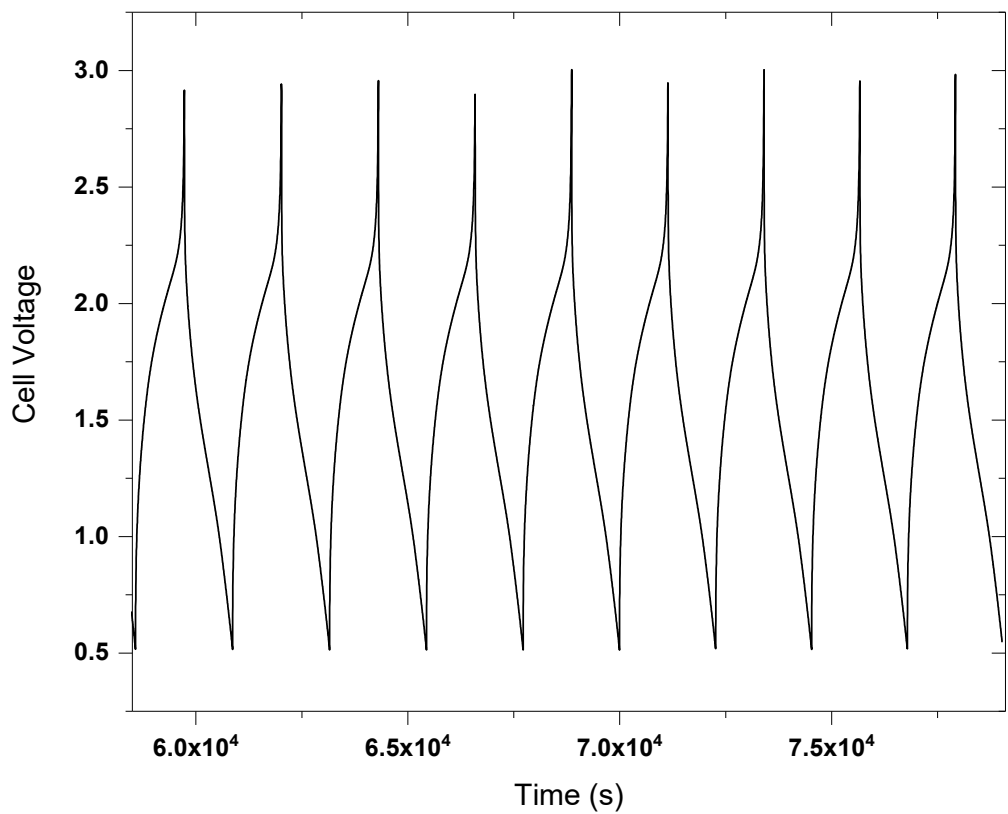

**Figure S3:** A snapshot of galvanostatic charge/discharge curves taken from an AEF 10 3D full cell with the exact configuration shown in Figure 5 in the main text. The applied current was 50 μA/cm$^2$. The cell had a solid electrolyte (ALD LPZ) only 40nm in thickness, which was fully conformal and provided full electrical isolation. This thickness of LPZ was the lower limit to achieve a reasonable (~50%) device yield per battery chip.

# Figure S3: XPS of the LiV$_2$O$_5$/LPZ Interface

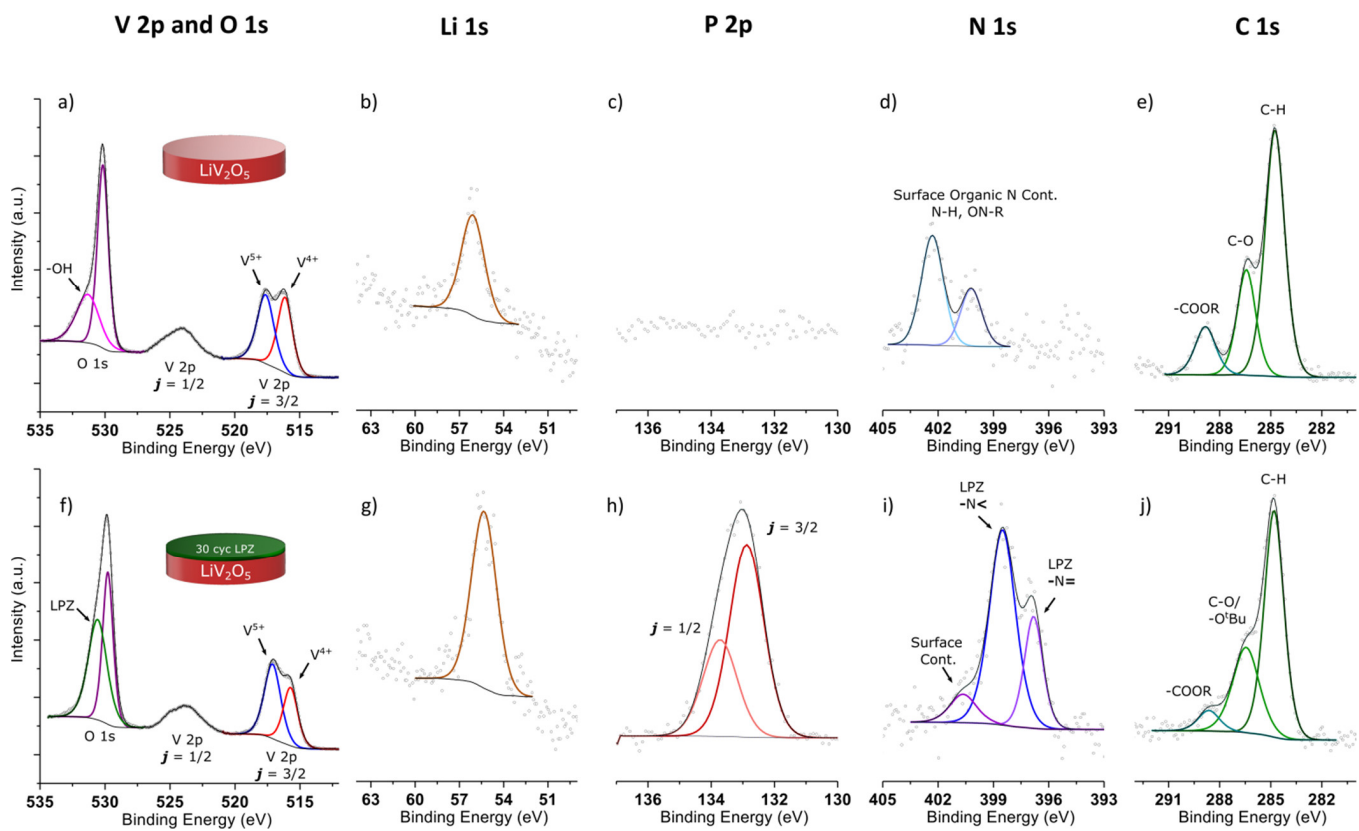

**Figure S4:** *In-situ* XPS characterization of the LiV$_2$O$_5$/LPZ interface. (a-e) High resolution component spectra of uncoated LiV$_2$O$_5$ taken after prelithiation. (f-j) High resolution spectra of the LiV$_2$O$_5$/LPZ interface following the application of 30 ALD LPZ cycles at 250C, resulting in the growth of ~1 nm of LPZ. The sample was transferred directly to the spectrometer from the ALD reactor under UHV conditions.

Figure S4 outlines the results of an experiment to characterize the interface chemistry at the LiV$_2$O$_5$/LPZ interface. An electrochemically lithiated LiV$_2$O$_5$ thin film was characterized via XPS, transferred under UHV to an ALD reactor in which it was exposed to 30 cycles of ALD LPZ [LiO$^t$Bu + Diethyl phosphoramidate (DEPA)] at 250C. The sample was then transferred back into the spectrometer without breaking vacuum. The ALD process grew a film with nominal thickness of 1.8nm, though the actual thickness is likely less than this due to a nucleation period of a few cycles. Photoelectrons from the LiV$_2$O$_5$ film are able to penetrate through this thickness of overlayer.

The primary features of the LiV$_2$O$_5$ surface include the presence of a slight solid-electrolyte interphase developed during electrochemical prelithiation, indicated by the presence of oxidized carbon species (Figure S4e). The O 1s peak (Figure S4a) shows a primary component at 530.2 eV associated with the oxide, as well as high binding energy shoulder associated with both the SEI and surface hydroxylation. The valence state of vanadium ions in the film can be determined by fitting the V 2p $j$=3/2 peak, which reveals an equal population of V$^{4+}$ (516.2 eV) and V$^{5+}$ (517.7 eV), as expected. There is a very small amount of organic nitrogen species initially present on the surface, whose origin is unknown.

After ALD LPZ deposition, XPS detects the presence of highly oxidized P, as well as an increase in the amount of detected Li and N, as expected. The average oxidation state of $LiV_2O_5$ surprisingly increases slightly, as the magnitude of the $V^{4+}$ component drops (Figure S4f). This rules out direct chemical lithiation of the $LiV_2O_5$ film via exposure to surface-adsorbed LiO$^t$Bu, in which case we would expect to see further reduction of the vanadium centers. We hypothesize that the extra oxidation at the surface could be due instead to DEPA directly reacting with Li ions in the underlying substrate during the formation of the first few monolayers of LPZ.

The N 1s spectrum also suggests that the first few monolayers of LPZ differ chemically from "bulk" LPZ (characterized below in Figure S5). The two characteristic peaks of ALD LPZ are detectable, which include the component associated with doubly-linked N (-N=) at 396.8 eV and one associated with triply bonded N (-N<) at 398.5 eV. However, the intensity ratio of this pair differs considerably from the bulk, in which the (-N=) component dominates (Figure S5b). One possible interpretation is that the LPZ directly at the $LiV_2O_5$/LPZ interface is more disordered due to interactions with both the cathode film and contaminant surface species. A similar effect has been previously observed for sputtered LiPON deposited on LCO, and may play an important role in understanding charge transfer resistance. [11]

# Figure S4: XPS of the LPZ/SnN$_x$ Interface

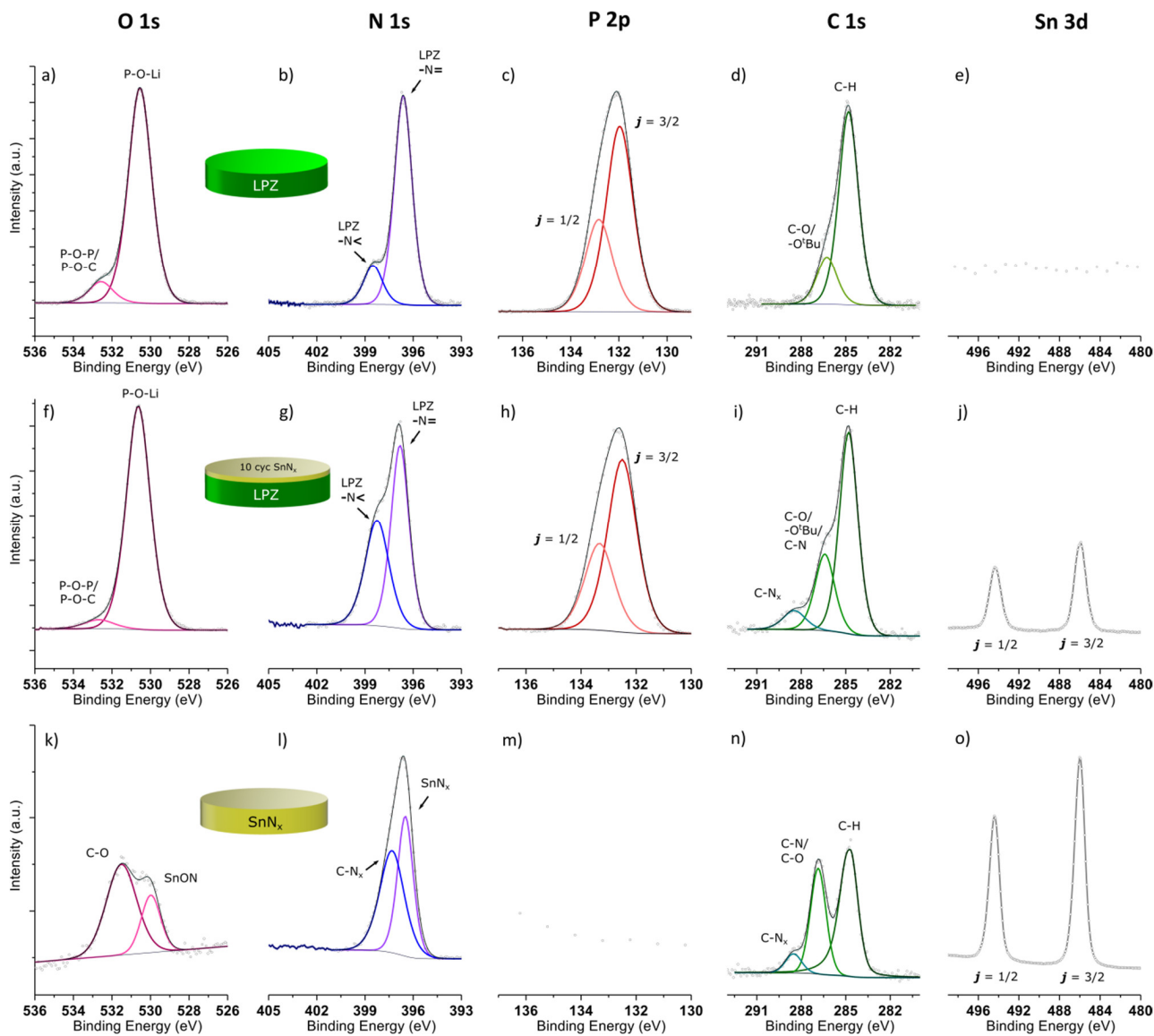

**Figure S5:** *In-situ* XPS characterization of the LPZ/SnN$_x$ interface. The Li 1s was omitted from this dataset as it did not shift or split in any way. (a-e) High resolution component spectra of pristine LPZ grown at 250C. (f-j) High resolution spectra of the LPZ/SnN$_x$ interface following the application of 10 ALD SnN$_x$ cycles at 200C, resulting in the growth of a maximum of ~0.5 nm of SnN$_x$. The total exposure time to N$_2$ plasma was 200s. (k-o) High resolution component spectra of thick (i.e. with no underlayer contribution) ALD SnN$_x$ grown at 200C. All samples were characterized without air exposure.

The increased impedance in all-ALD full cells (Figure 6d in the main text) led us to characterize the ALD LPZ/SnN$_x$ interface directly using the same general procedure as in Figure S4. Figures S5a-e show high resolution XPS core level spectra from a LPZ film grown at 250C, whose components are analyzed in detail in a previous publication.[2] The typical stoichiometry of LPZ films grown at 250C is Li$_{1.7}$PO$_{2.1}$N (plus residual hydrocarbons). We were interested in observing whether or not exposure of this surface to the SnN$_x$ process precursors (TDMASn and a remote N$_2$ plasma) resulted in detectable chemical decomposition of the LPZ which could explain the increase in cell impedance in the full cell vs. half-cells.

Panels f-j show the surface chemistry after 10 cycles of the SnN$_x$ ALD process at 200C. This results in the accumulation of only approx. 2 atomic % Sn by XPS quantification, and so the vast majority of photoelectron intensity measured for the O 1s, N 1s, and C 1s lines still originate from the LPZ. There are relatively few differences in the underlying LPZ surface chemistry, but a decrease in the intensity of the bridging oxygen component of the O 1s at 532.7eV (Figure S5f) and an increase in the triply bonded nitrogen component at 398.2 eV (Figure S5g) suggest a possible reorganization of the LPZ at the anode/electrolyte interface. However, there is no strong evidence to suggest that this is responsible for the increase in cell impedance observed, especially given that there are many examples of conductive LiPON-family films with N 1s spectra closely resembling that measured here. There is no evidence for either the oxidation or reduction of P. The binding energy of Sn atoms at the interface (486 eV) matches that of the "bulk" film exactly, suggesting as well that the TDMASn precursor does not interact with the LPZ substrate in any unexpected ways.

In summary, XPS of the SnN$_x$/LPZ interface indicates a few minor changes in the LPZ chemistry, but the electrolyte overall tolerates the overgrowth of the anode well. While this leaves the additional impedance unresolved (the next step is to carefully examine effects from ion milling such as local heating), it bodes well for the compatibility of LPZ with CVD or ALD of other anode materials in the future.

**Figure S5: Cross Section of Full Cell Stack Characterized in Figure 7**

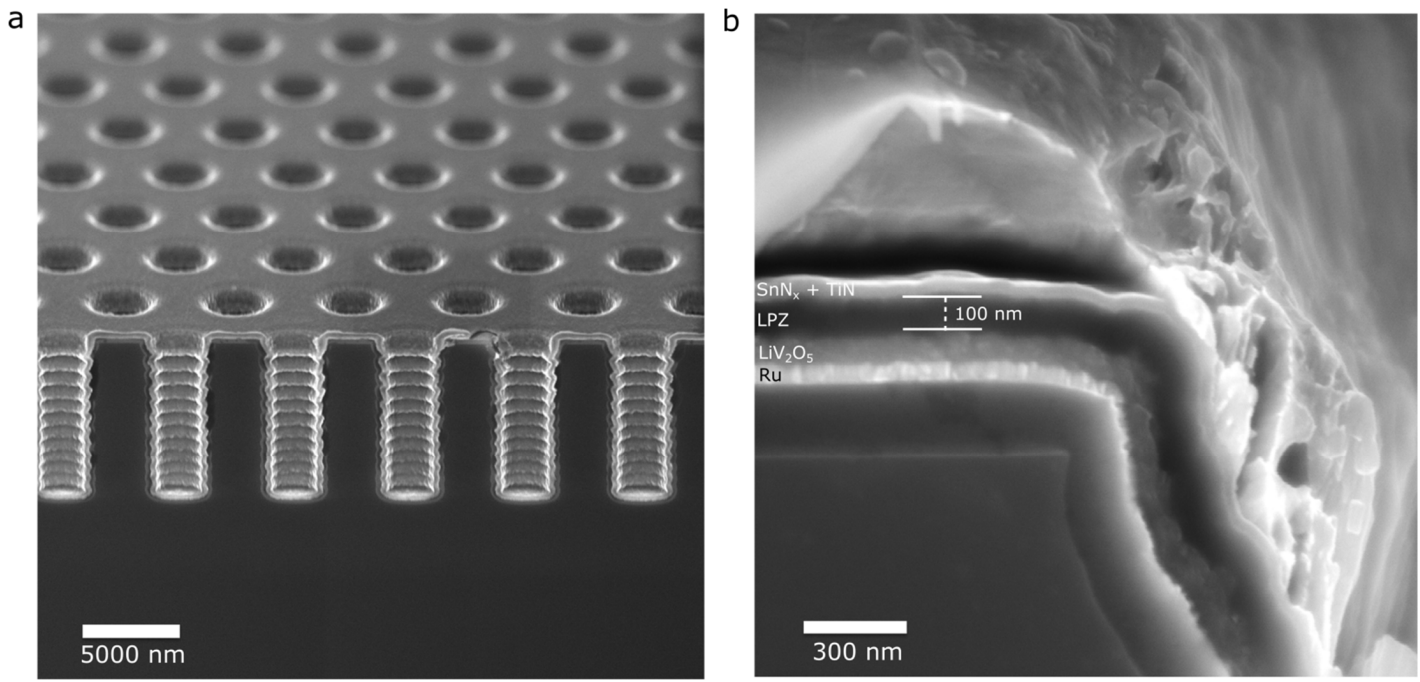

**Figure S6:** SEM characterization of an AEF 4 3D full cell from the same deposition run which produced the film stack characterized in cells of differing AEF in Figure 7 of the main text. (a) Tilted view of the 3D array with full cell deposited (from a region without the Cu capping layer) (b) Close-in view of the film stack taken showing film thicknesses.

# Figure S6: Discharge Voltage Decay in 3D Full Cells

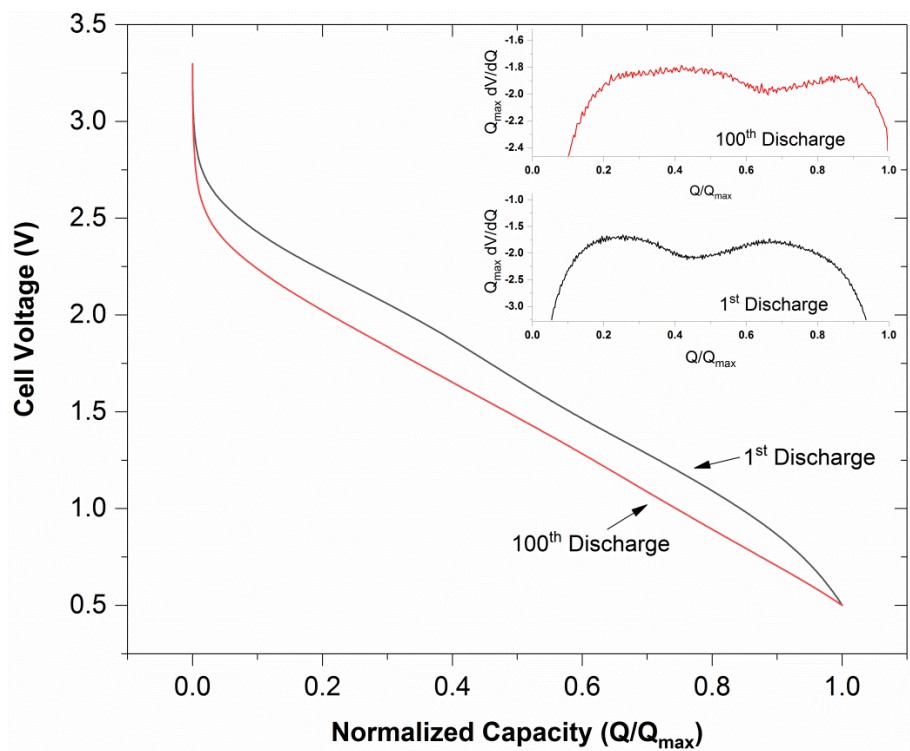

**Figure S6:** A comparison of the discharge curves for the first and 100[th] cycles of the AEF 10 cell characterized in Figure 7 of the main text. The capacity is normalized to the measured discharge capacity at the cutoff potential $Q_{max}$ to illustrate the decrease in discharge potential of approx. 0.2V, likely due to free lithium loss from the cell. Losing free lithium results in underutilization of the anode, causing its average potential to increase vs. Li/Li[+]. For both curves, two phase transitions (indicated by local maxima) associated with the two-step lithium insertion reaction of $LiV_2O_5$ can still be detected from the derivative dQ/dV, plotted in the inset.